\documentstyle[prd,aps,psfig,amstex,floats,preprint,tighten]{revtex}

\newcommand{\ifm}[1]{\relax\ifmmode #1\else $#1$\fi}
\renewcommand{\d}{\mbox{d}}
\newcommand{\etal}{\mbox{\it et al.}}
\newcommand{\ie}{\mbox{\it i.e.}}
\newcommand{\eg}{\mbox{\it e.g.}}

\newcommand{\mevcc}{\ifm{\mbox{MeV}}}
\newcommand{\gevcc}{\ifm{\mbox{GeV}}}
\newcommand{\gevc}{\ifm{\mbox{GeV}}}
\newcommand{\gev}{\ifm{\mbox{GeV}}}
\renewcommand{\l}{\ifm{\ell}}
\newcommand{\xbar}{\ifm{\overline{x}}}
\newcommand{\nubar}{\ifm{\overline{\nu}}}
\newcommand{\pbar}{\ifm{\overline{p}}}
\newcommand{\tbar}{\ifm{\overline{t}}}
\newcommand{\bbar}{\ifm{\overline{b}}}
\newcommand{\qbar}{\ifm{\overline{q}}}
\newcommand{\ppbar}{\ifm{p\pbar}}
\newcommand{\ttbar}{\ifm{t\tbar}}

\newcommand{\qqbar}{\ifm{q\qbar}}
\newcommand{\px}{\ifm{p_x}}
\newcommand{\py}{\ifm{p_y}}

\newcommand{\pt}{\ifm{p_T}}
\newcommand{\ptv}{\ifm{\vec\pt}}
\newcommand{\mpt}{\mbox{$\rlap{\kern0.1em/}\pt$}}
\newcommand{\mpx}{\mbox{$\rlap{\kern0.1em/}\px$}}
\newcommand{\mpy}{\mbox{$\rlap{\kern0.1em/}\py$}}
\newcommand{\mptv}{\mbox{$\rlap{\kern0.1em/}\vec\pt$}}
\newcommand{\NMC}{N^{\rm MC}}
\newcommand{\PM}{\ifm{\pm}}
\newcommand{\lt}{\ifm{<}}
\newcommand{\gt}{\ifm{>}}

\newcommand{\me}{\mbox{$\rlap{\kern0.1em/}p$}}

\renewcommand{\o} {\ifm{\lbrace o \rbrace}}

\renewcommand{\v} {\ifm{\lbrace v \rbrace}}
\renewcommand{\P}{\ifm{{\cal P}}}

\newcommand{\M}{\ifm{{\cal M}}}
\newcommand{\MWT}{\M WT}
\newcommand{\vWT}{$\nu$WT}

\newcommand{\Q}{\ifm{{\cal Q}}}
\renewcommand{\L}{\ifm{{L}}}
\newcommand{\ipb}{{\rm pb}^{-1}}

\newcommand{\mW}{M_W}
\newcommand{\PRL}{Phys. Rev. Lett.}

\newcommand{\PR}{Phys. Rev.}
\newcommand{\NP}{Nucl. Phys.}
\newcommand{\NIM}{Nucl. Instrum. Methods in Phys. Res.}

\newcommand{\GEANT}{{\sc geant}}
\newcommand{\ISAJET}{{\sc isajet}}
\newcommand{\HERWIG}{{\sc herwig}}

\begin{document}
\pagestyle{myheadings}
\onecolumn
\preprint{FERMILAB-Pub-98-261-E}

\title {Measurement of the Top Quark Mass in the Dilepton Channel}

\author{
\centerline{The D\O\ Collaboration\thanks{Authors listed on the following page.
            \hfill\break 
            Submitted to Physical Review D.}}
}

\address{
\centerline{Fermi National Accelerator Laboratory, Batavia, Illinois 60510}
}

\date{August 26, 1998}

\maketitle

\vspace{0.5in}
\begin{abstract}
We report a measurement of the top quark mass using six candidate events
for the process $\ppbar\to\ttbar+X\to\l^+\nu b\l^-\nubar\bbar+X$, observed
in the D\O\ experiment at the Fermilab \ppbar\ collider. Using maximum
likelihood fits to the dynamics of the decays, we measure a mass for the top
quark of
$m_t = 168.4 \pm 12.3 \mbox{ (stat)} \pm 3.6 \mbox{ (syst)}$ \gevcc.
We combine this result with our previous measurement in the
$\ttbar\to\l+\mbox{jets}$ channel
to obtain $m_t = 172.1 \pm 7.1$ \gevcc\ as the best
value of the mass of the top quark measured by D\O.
\end{abstract}


\vskip 1cm
%
\begin{center}
B.~Abbott,$^{40}$                                                             
M.~Abolins,$^{37}$                                                            
V.~Abramov,$^{15}$                                                            
B.S.~Acharya,$^{8}$                                                           
I.~Adam,$^{39}$                                                               
D.L.~Adams,$^{48}$                                                            
M.~Adams,$^{24}$                                                              
S.~Ahn,$^{23}$                                                                
H.~Aihara,$^{17}$                                                             
G.A.~Alves,$^{2}$
% ...
\end{center}
\vskip 0.50cm
\normalsize
\centerline{(D\O\ Collaboration)}

\small
\it
\centerline{$^{1}$Universidad de Buenos Aires, Buenos Aires, Argentina}       
\centerline{$^{2}$LAFEX, Centro Brasileiro de Pesquisas F{\'\i}sicas,         
                  Rio de Janeiro, Brazil}                                     
\centerline{$^{3}$Universidade do Estado do Rio de Janeiro,                   
                  Rio de Janeiro, Brazil}                                     
\centerline{$^{4}$Universidad de los Andes, Bogot\'{a}, Colombia}             
\centerline{$^{5}$DAPNIA/Service de Physique des Particules, CEA, Saclay,     
                  France}                                                     
\centerline{$^{6}$Panjab University, Chandigarh, India}                       
\centerline{$^{7}$Delhi University, Delhi, India}                             
\centerline{$^{8}$Tata Institute of Fundamental Research, Mumbai, India}      
\centerline{$^{9}$Kyungsung University, Pusan, Korea}                         
\centerline{$^{10}$Seoul National University, Seoul, Korea}                   
\centerline{$^{11}$CINVESTAV, Mexico City, Mexico}                            
\centerline{$^{12}$Institute of Nuclear Physics, Krak\'ow, Poland}            
\centerline{$^{13}$Institute for Theoretical and Experimental Physics,        
                   Moscow, Russia}                                            
\centerline{$^{14}$Moscow State University, Moscow, Russia}                   
\centerline{$^{15}$Institute for High Energy Physics, Protvino, Russia}       
\centerline{$^{16}$University of Arizona, Tucson, Arizona 85721}              
\centerline{$^{17}$Lawrence Berkeley National Laboratory and University of    
                   California, Berkeley, California 94720}                    
\centerline{$^{18}$University of California, Davis, California 95616}         
\centerline{$^{19}$University of California, Irvine, California 92697}        
\centerline{$^{20}$University of California, Riverside, California 92521}     
\centerline{$^{21}$Florida State University, Tallahassee, Florida 32306}      
\centerline{$^{22}$University of Hawaii, Honolulu, Hawaii 96822}              
\centerline{$^{23}$Fermi National Accelerator Laboratory, Batavia,            
                   Illinois 60510}                                            
\centerline{$^{24}$University of Illinois at Chicago, Chicago,                
                   Illinois 60607}                                            
\centerline{$^{25}$Northern Illinois University, DeKalb, Illinois 60115}      
\centerline{$^{26}$Northwestern University, Evanston, Illinois 60208}         
\centerline{$^{27}$Indiana University, Bloomington, Indiana 47405}            
\centerline{$^{28}$University of Notre Dame, Notre Dame, Indiana 46556}       
\centerline{$^{29}$Purdue University, West Lafayette, Indiana 47907}          
\centerline{$^{30}$Iowa State University, Ames, Iowa 50011}                   
\centerline{$^{31}$University of Kansas, Lawrence, Kansas 66045}              
\centerline{$^{32}$Louisiana Tech University, Ruston, Louisiana 71272}        
\centerline{$^{33}$University of Maryland, College Park, Maryland 20742}      
\centerline{$^{34}$Boston University, Boston, Massachusetts 02215}            
\centerline{$^{35}$Northeastern University, Boston, Massachusetts 02115}      
\centerline{$^{36}$University of Michigan, Ann Arbor, Michigan 48109}         
\centerline{$^{37}$Michigan State University, East Lansing, Michigan 48824}   
\centerline{$^{38}$University of Nebraska, Lincoln, Nebraska 68588}           
\centerline{$^{39}$Columbia University, New York, New York 10027}             
\centerline{$^{40}$New York University, New York, New York 10003}             
\centerline{$^{41}$University of Rochester, Rochester, New York 14627}        
\centerline{$^{42}$State University of New York, Stony Brook,                 
                   New York 11794}                                            
\centerline{$^{43}$Brookhaven National Laboratory, Upton, New York 11973}     
\centerline{$^{44}$University of Oklahoma, Norman, Oklahoma 73019}            
\centerline{$^{45}$Brown University, Providence, Rhode Island 02912}          
\centerline{$^{46}$University of Texas, Arlington, Texas 76019}               
\centerline{$^{47}$Texas A\&M University, College Station, Texas 77843}       
\centerline{$^{48}$Rice University, Houston, Texas 77005}                     
\vfill\eject
\normalsize

\newpage
\tableofcontents
\newpage

\section{ Introduction }
\label{sec-intro}
The mass of  the top  quark is a free  parameter in  the  standard model of the
electroweak interactions  \cite{SM}. It arises from  the Yukawa coupling of the
top quark to the Higgs field, which is  not constrained by the model. Through
radiative corrections, the value of the top quark mass affects predictions of
the  standard  model for many processes. For example, the prediction for the
mass of the $W$ boson varies by approximately 7 \mevcc\footnote{We use natural
units with $\hbar=c=1$.} for every 1 \gevcc\
change in the mass of the top quark \cite{mw_v_mt}. Precise measurements of the
masses of the top quark and the $W$ boson constrain the mass of the Higgs
boson. This dependence can be turned around and the top
quark mass predicted from measurements of  electroweak processes within
the framework of the  standard model. Such an  analysis gives $158^{+14}_{-11}$
\gevcc\ for the top quark mass  \cite{ewwg}. In this sense, a measurement of the
top quark mass constitutes a consistency test of the standard model prediction.

The top quark is  the only fermion with  a mass close to the
vacuum  expectation value of the  Higgs field,  or equivalently, with a Yukawa
coupling close to unity. It is therefore possible that by studying the
properties  of the  top  quark we can learn  more about   electroweak symmetry
breaking.

The  Fermilab   Tevatron  produces top  quarks  in  collisions  of  protons and
antiprotons at $\sqrt{s}=1.8$ TeV. The Tevatron provided the first experimental
confirmation of the existence of the top quark \cite{top_discovery}. In \ppbar\
collisions top quarks are produced predominantly in \ttbar\ pairs. The standard
model predicts the top quark primarily ($\gt99$\%) to decay to $Wb$. The decay
modes of the $W$ boson then define the signatures of \ttbar\  decays.  If both
$W$ bosons  decay  leptonically the signature contains two charged leptons with
high  \pt. We call this the dilepton channel. Events in which one of the  $W$
bosons decays   leptonically and the  other into  jets contain one high \pt\
charged lepton and high \pt\  hadron jets. We call this the
lepton+jets channel. In the all-jets channel both $W$ bosons decay into  jets.

The D\O\ collaboration was first to measure the mass of the top quark in the
dilepton channel \cite{D0_mtop_ll_PRL,Varnes_thesis}. In this
article we present  a more detailed  account of this analysis. The most precise
measurements of the top quark mass have been obtained using the lepton+jets
channel \cite{D0_mtop_lj,CDF_mtop_lj}. Table \ref{tab:mtpub} lists  previously
published measurements of the top quark mass.

\begin{table}[h]
\begin{center}
\caption{Published measurements of the top quark mass. The first uncertainty is
statistical, the second systematic.}
\begin{tabular}{lcr@{}r@{\PM}r@{}r@{\PM}r@{}r}
Experiment      & Channel   & \multicolumn{6}{c}{Mass}              \\
\hline
D\O \protect\cite{D0_mtop_lj}     & lepton+jets & 173&.3& 5&.6&5&.5 \gevcc \\
D\O \protect\cite{D0_mtop_ll_PRL} & dilepton    & 168&.4&12&.3&3&.6 \gevcc \\
CDF \protect\cite{CDF_mtop_lj}    & lepton+jets & 175&.9& 4&.8&4&.9 \gevcc \\
CDF \protect\cite{CDF_mtop_ll}    & dilepton    & 161&  &17&  &10&  \gevcc \\
CDF \protect\cite{CDF_mtop_jj}    & all-jets    & 186&  &10&  &12&  \gevcc \\
\end{tabular}
\label{tab:mtpub}
\end{center}
\end{table}

The measurement described in this paper is based on an integrated luminosity of
approximately $125$ $\ipb$, recorded by the D\O\ detector during the 1992--1996
collider runs. We first give a brief description of the experimental setup
(Sect. \ref{sec-exp}), data reconstruction (Sect. \ref{sec-pid}) and
calibration procedures (Sect. \ref{sec-escale}). We then describe the
selection of the event sample (Sect. \ref{sec-evt}), the mass analysis of the
selected events (Sect. \ref{sec-mt}), the maximum likelihood fit to the data
(Sect. \ref{sec-fit}), and the systematic uncertainties associated with the
fit (Sect. \ref{sec-syst}). Finally we summarize the results and combine them
with the measurement in the lepton+jets channel (Sect. \ref{sec-results}).

\section{ Detector }
\label {sec-exp}
\label{d0detector}

D\O\ is a multipurpose detector designed to study \ppbar\ collisions at
high energies. The detector was commissioned at the Fermilab Tevatron
during  the summer of 1992. A full description of the detector can be found in
Ref. \cite{d0nim}. Here, we describe only briefly the properties of the detector
that are relevant for the mass measurement in the dilepton channel.

We specify detector coordinates in a system with its origin defined by
the center of the detector and the $z$-axis defined by the proton beam.
The $x$-axis points out of the Tevatron ring and the $y$-axis up.
We use $\phi$ to denote the azimuthal coordinate and $\theta$ for the polar
angle. Rather than $\theta$, we often use the pseudorapidity 
$\eta = \tanh^{-1}(\cos\theta)$.

\begin{figure}[htpb]
\psfig{figure=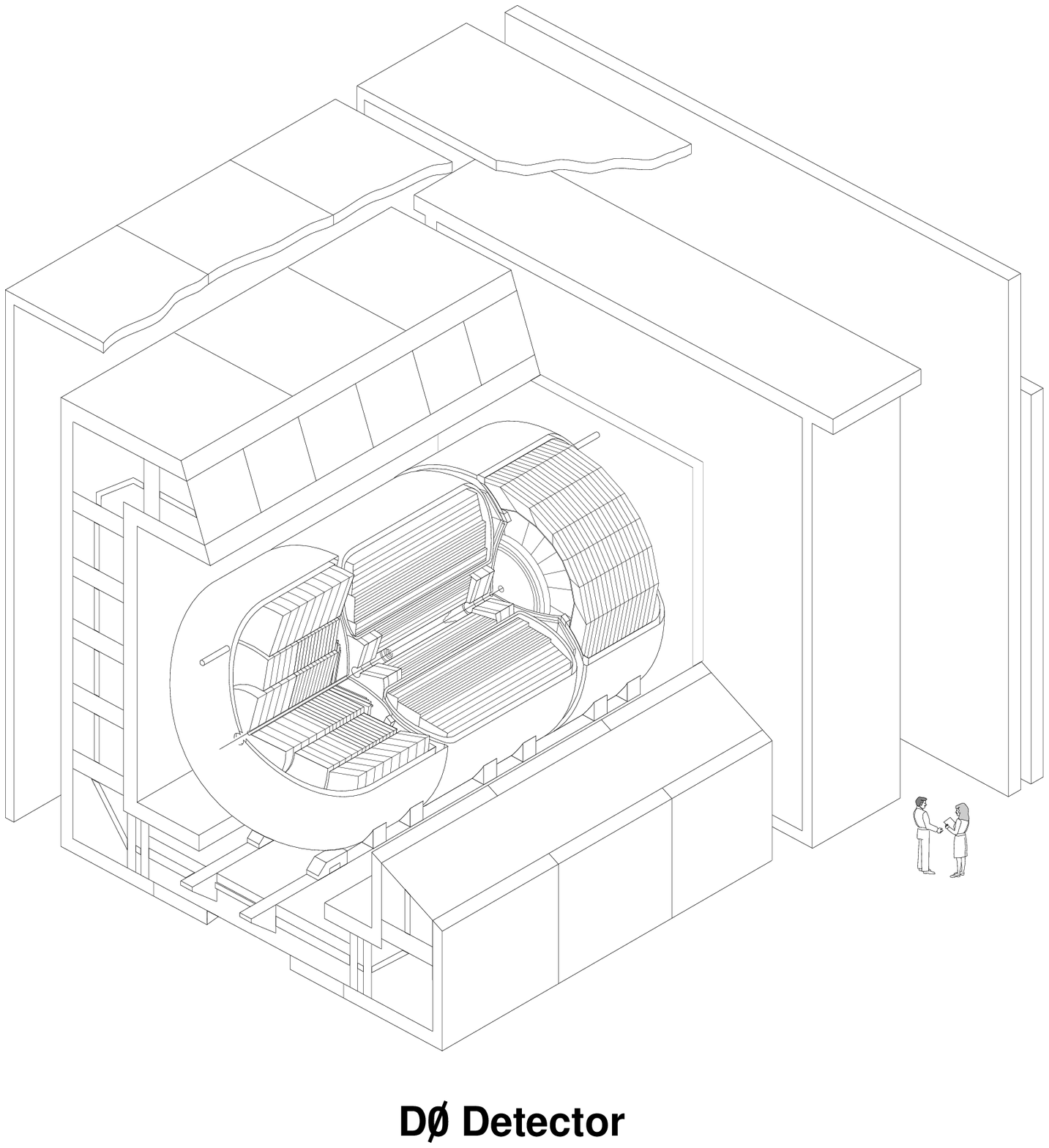,width=\hsize}
\caption{Cut away isometric view of the D\O\ detector.}
\label{fig:figdet1}
\end{figure}

The detector consists of three primary systems: central tracking,
calorimeter, and muon spectrometer. A
cut away view of the detector is shown in Fig.~\ref{fig:figdet1}.

The nonmagnetic central tracking system consists of four subdetectors that
measure the trajectories of charged particles: a vertex drift chamber, a
transition radiation detector, a central drift chamber, and two forward drift
chambers. These chambers also measure ionization to identify tracks
from single charged particles and $e^+e^-$ pairs from photon conversions.  
The central tracking system covers the region $|\eta| < 3.2$.

The uranium-liquid argon calorimeter is divided into three parts, the central
calorimeter and the two end calorimeters, and covers the pseudorapidity range
$|\eta| < 4.2$.  Longitudinally, the calorimeter is segmented into an
electromagnetic (EM) section with fine sampling and a hadronic section with
coarser sampling. The calorimeter is segmented transversely into quasiprojective
towers with $\Delta\eta\times\Delta\phi$ = $0.1\times 0.1$. The third layer of
the electromagnetic calorimeter, where EM showers are expected to peak, is
segmented twice as finely in each direction. The hadronic calorimeter modules
back up any cracks in the coverage of the EM calorimeter modules such that there
are no projective cracks in the calorimeter, ensuring good resolution for
the measurement of transverse momentum balance.

Since muons from top quark decays predominantly populate the central region, we
use only the central portion of the muon system, which covers $|\eta| < 1.7$.
This system consists of four planes of proportional drift tubes in front of
magnetized iron toroids with a magnetic field of 1.9~T and two groups of three
planes of proportional drift tubes behind the toroids. The magnetic field lines
and the wires in the drift tubes are oriented transversely to the beam
direction. The momentum is obtained from the deflection of the muon in the
magnetic field of the toroid.

\section{ Particle Identification }
\label {sec-pid}
The particle identification algorithms used for electrons, muons, and jets are
the same as in previously published analyses\cite{xsec_PRL}. 
We summarize them in the following sections.

\subsection{Electrons}

Electron candidates are first identified by finding isolated clusters
of energy in the EM calorimeter along with a matching track in the central
detector. We accept electron candidates with $|\eta| \le 2.5$.
Final identification is based on a likelihood test on the following
five variables:
\begin{itemize}
\item The agreement of the shower shape with the expected shape of an
      electromagnetic shower, computed using the full covariance matrix of the
      energy depositions in the cells of the electromagnetic calorimeter.
\item The electromagnetic energy fraction, defined as the ratio of the
      shower energy found in the electromagnetic calorimeter
      to the total shower energy.
\item A measure of the distance between the track and the cluster
      centroid.
\item The ionization $dE/dx$ along the track.

\item A variable characterizing the energy deposited in the
	transition radiation
      detector.

\end{itemize}
To a good approximation, these five variables are
independent of each other for electron showers.

Electrons from $W$~boson decay tend to be isolated.  Thus, we make the
additional cut
\begin{equation}
{E_{\text{tot}}(0.4) - E_{\text{EM}}(0.2) \over E_{\text{EM}}(0.2)} < 0.1,
\end{equation}
where $E_{\text{tot}}(0.4)$ is
the energy within $\Delta R < 0.4$ of the cluster centroid and
$E_{\text{EM}}(0.2)$ is the energy in the EM calorimeter within
$\Delta R < 0.2$.  $\Delta R$ is defined as $\sqrt{\Delta\eta^2 +
\Delta\phi^2}$.

\subsection{Muons}

Two types of muon selection are used in this analysis. The first is used to
identify isolated muons from $W\rightarrow \mu\nu$ decay.
The second type of muon selection is used to tag $b$-jets by
identifying muons consistent with
originating from $b\rightarrow \mu + X$ decay.
We accept muons with $|\eta|< 1.7$. 
Besides cuts on the muon track quality,
both selections require
that the energy deposited in the calorimeter along a muon track
be at least that expected from a minimum ionizing particle. For isolated muons,
such as those from $W$ boson decays, we require $\Delta R_{\mu,j} > 0.5$ for
the distance $\Delta R_{\mu,j}$ in the $\eta - \phi$ plane between the muon and
any jet. For soft muons in jets, such as those from $b \rightarrow \mu + X$ decay,
we require $\pt \ge 4\ \gevc$ and $\Delta R_{\mu,j} < 0.5$.
The efficiency$\times$acceptance for either muon selection with these cuts is 
about $64\%$.

\subsection{Jets}

Jets are reconstructed in the calorimeter using a fixed-size cone
algorithm.  We use a cone size of $\Delta R = 0.5$.
See Ref. \cite{jetreco} for a detailed
description of the jet reconstruction algorithm.

\subsection{Missing Transverse Momentum}

The missing transverse momentum, $\mptv$, is the momentum required to balance
the measured momenta in the event ($\sum \ptv + \mptv = 0$).  
In the calorimeter, we calculate $\mptv$ as
\begin{equation}
\mptv^{\rm cal} = - \sum_i E_i\sin \theta_i
\left(\begin{array}{c}\cos\phi_i\\ \sin\phi_i\end{array}\right),
\end{equation}
where $i$ runs over all calorimeter cells, $E_i$ is the energy
deposited in the $i^{th}$ cell, and $\phi_i$ is the azimuthal and $\theta_i$ 
the polar angle of the
$i^{th}$ cell. When there are muons present in the event we refine
the calculation 
\begin{equation}
\mptv = \mptv^{\rm cal} - \sum_k \ptv^{\,\mu_k},
\end{equation}
where $\ptv^{\,\mu}$ is the transverse momentum of the muon as measured by
the muon system.

\section { Energy Scale Calibration }
\label{sec-escale}
\subsection{ Electron Energy Scale }

The measurement of the energy $E$ of electromagnetic showers in the calorimeter
is calibrated using $Z\to ee$, $J/\psi\to ee$, and $\pi^0\to \gamma\gamma$
decays to a precision of 0.08\% at $E=M_Z/2$ and to 0.6\% at $E=20$ GeV
\cite{W_mass}. The electron energy scale calibration therefore does not give
rise to any significant uncertainty in the top quark mass measurement.

\subsection { Muon Momentum Scale }

The muon momentum scale, calibrated using $J/\psi \rightarrow \mu\mu$ and
$Z \rightarrow \mu\mu$ candidates, has an uncertainty of 2.5\%. Its effect
on our measurement of the top quark mass was determined by varying
the muon momentum scale in Monte Carlo samples of $t \bar t$ events with
$m_t=170$ \gevcc. The tests
indicate that the relation between muon scale and top quark mass error
is given by
\begin{equation}
    \delta m_t = 12\ \gevcc\, {\delta p_T^{\mu}\over p_T^{\mu}}.
\end{equation}

Hence, the 2.5\% uncertainty in muon momentum scale leads to a systematic
uncertainty of 0.3 \gevcc\ in our measurement of the top quark mass. This
uncertainty is completely negligible compared to the effect of the jet energy
scale.

\subsection { Jet Energy Scale }
\label{sec-jscale}

The jet energy scale is calibrated relative to the electromagnetic energy scale
by balancing the transverse momentum in events with jets and electromagnetic
showers \cite{jet_energy_scale}.
The exercise is carried out separately and symmetrically for both data and
Monte Carlo.

In addition to the corrections in Ref. \cite{jet_energy_scale} we apply
an $\eta$-dependent correction derived from a comparison
between $\gamma$+jet events in data and Monte Carlo events created using 
the \HERWIG \cite{herwig} event generator and a \GEANT\ \cite{geant} based
detector simulation.
We also correct jets that contain a muon, indicative of
a semileptonic $b$ quark decay, to 
compensate on average for the energy carried away by the undetected
neutrino.
These corrections are identical to those used and detailed in the mass
analysis based on the lepton+jets final states \cite{D0_mtop_lj}
with the exception that no attempt is made
to account for gluon radiation outside of the jet cone.
Rather, the procedure in the dilepton analysis is to explicitly account for
additional reconstructed jets, as described in Sect.~\ref{sec-jet_comb}.

We estimate the degree of possible residual discrepancy between the
jet energy response of the detector and the Monte Carlo simulation from
the energy balance between electromagnetic energy clusters and jets from collider
data, compared to photon+jets Monte Carlo, as a function of photon \pt.
The data constrain the possible mismatch to less than
$\pm(2.5\%+0.5\ \gev)$ in the jet energy \cite{D0_mtop_lj}. This uncertainty
gives rise to a significant systematic uncertainty in our top quark mass
measurement (see Sect. \ref{sec-syst_jscale}).

\section{ Event Selection }
\label{sec-evt}
\subsection {Basic Event Selection Criteria}

The event selection for the dilepton mass analysis is almost identical to that
used for the measurement of the cross section \cite{xsec_PRL}. 
We require two charged leptons
($e$, $\mu$) and at least two jets in the events. In addition we cut on 
global event quantities like $\mpt$ and $H_T$. The basic kinematic selection
criteria are summarized in Table \ref{tab:dilep_cuts}. The variable $H_T$ is
defined as
\begin{equation}
H_T =
\begin{cases}
\sum p_T^j + p_T^{e_1}& \text{for the $ee$ and $e\mu$ channels}; \\
\sum p_T^j            & \text{for the $\mu\mu$ channel},
\end{cases}
\end{equation}
where $e_1$ is the leading electron in $ee$ events. The sum is over all
jets with \pt $> 15$ \gevc\ and $|\eta| < 2.5$.  Muons are not included in the
sum because their momenta are measured less precisely. $H_T$ gives
good rejection against background processes, which typically have less jet
activity along with the dilepton signature. 

\begin{table}[h]
\begin{center}
\caption{Kinematic and fiducial cuts used in selecting dilepton events.
\label{tab:dilep_cuts}}
\begin{tabular}{lcr@{ }lr@{ }lr@{ }l}
Objects       &                  & \multicolumn{2}{c}{$ee$} & 
\multicolumn{2}{c}{$e\mu$} & \multicolumn{2}{c}{$\mu\mu$}    \\
\hline
2 Leptons     & $\pt^\l$         & $>20$ &\gevc  & $>15$ &\gevc  & $>15$ &\gevc \\
              & $|\eta^\l|$      & $<2.5$ &      & $<1.7$ &      & $<1.7$ &     \\ 
$\geq 2$ Jets & $\pt^j$          & $>20$ &\gevc  & $>20$ &\gevc  & $>20$ &\gevc \\
              & $|\eta^j|$       & $<2.5$ &      & $<2.5$ &      & $<2.5$ &     \\
Event         & \mpt             & --- &         & $>10$ &\gevc  & --- &        \\
              & $\mpt^{\rm cal}$ & $>25$ &\gevc  & $>20$ &\gevc  & --- &        \\ 
              & $H_T$            & $>120$ &\gevc & $>120$ &\gevc & $>100$ &\gevc \\
\end{tabular}
\end{center}
\end{table}

The event selection criteria are designed to identify events with two charged leptons 
and additional jets in the final state as expected from $\ttbar\to\l\l+X$
decays. The background in the $ee$ and $\mu\mu$ channels is dominated by $Z\to
ee$ and $Z\to\mu\mu$ decays. We apply additional criteria, described in the
following sections, that remove these particular backgrounds. Table 
\ref{tab:dilep_bkg} gives the number of background events expected in each
dilepton channel after all selection criteria are applied. Instrumental
backgrounds arise from particle misidentification, \eg\ mistaking a jet for an
electron.

\begin{table}[h]
\begin{center}
\caption{Expected numbers of background events.\label{tab:dilep_bkg}}
\begin{tabular}{cccc}
Background Source     & $ee$            & $e\mu$          & $\mu\mu$        \\ 
\hline
$Z\to\l\l$            & $0.058\pm0.012$ &   ---           &  $0.558\pm0.21$  \\
$Z\to\tau\tau\to\l\l$ & $0.078\pm0.022$ & $0.099\pm0.076$ &  $0.029\pm0.017$  \\
$WW$                  & $0.083\pm0.023$ & $0.074\pm0.018$ &  $0.007\pm0.004$ \\ 
Drell-Yan             & $0.054\pm0.030$ & $0.002\pm0.003$ &  $0.066\pm0.035$  \\ 
$\ttbar\to e+\hbox{jets}$ & 0.04        &  ---            & ---             \\
Instrumental          & $0.197\pm0.046$ &  $0.035\pm0.13$  &  $0.068\pm0.010$  \\ 
\hline
Total Background      & $0.51\pm0.09$   & $0.21\pm0.16$   & $0.73\pm0.25$  \\
\end{tabular}
\end{center}
\end{table}

\subsection {$e\mu$ Channel}
The $e\mu$ channel is the most powerful dilepton channel with twice the
branching ratio of the $ee$ and $\mu\mu$ channels and without the background
from $Z\to ee$ or $Z\to\mu\mu$ decays. The largest background is
$Z\to\tau\tau\to e\mu+X$,  which is suppressed by both  branching ratio and
kinematics. Instrumental backgrounds arise from $W$ bosons that decay to
$\mu\nu$ which are produced in association with jets, one of which is mistaken
for an electron.

We observe three events in this channel.

\subsection {$ee$ Channel}

The primary source of physics background in the $ee$ channel is $Z$ boson
production with associated jets. These events have no neutrinos and can be
rejected effectively by cutting on \mpt. We therefore require $\mpt>40$ \gevc\
if the dielectron invariant mass is within 12 \gevcc\ of the $Z$ boson mass 
peak.
Instrumental backgrounds arise from $W$+jets production or multijet events in
which jets fake the electron signature.

In this channel we extend our event selection criteria to include an additional
event that was not part of the final sample for the measurement of the cross 
section. This event passes all selection criteria, except that one of the
electron candidates has no matching track. This cluster is nevertheless
consistent with originating from an electron because the trajectory connecting
the vertex with
the cluster passes only through the two inner layers of the CDC. The inner two
layers do indeed have hits but to reconstruct a track, hits are required in at
least three layers. The lack of a reconstructed track could indicate a higher
probability for this electron to be misidentified. On the other hand one of the
jets contains a muon, which passes all requirements for the muon-tag analyses
reported in reference \cite{xsec_PRL}. 
A muon tag indicates that the jet
probably originates from the fragmentation of a $b$ quark. The probability of
tagging a jet from the fragmentation of a light quark or a gluon is quite
small. 
The presence of a $b$ jet reduces the likelihood that this event arises from
instrumental background sources and we therefore include it in the event
sample for the mass analysis.

We revise the background estimate for the $ee$ channel to include an additional
component due to the inclusion of this event. We compute the number of
additional background events expected if events are admitted that are missing a
matched track for one of the two electron candidates but have a muon tag.
In our data we find 11 events with one
electron candidate and three jets, one with muon tag.
In these events, there are 22 jets that could fake a second electron.
The probability for any one of these jets to mimic an electron signature without
matched track requirement is $8 \times 10^{-4}$\cite{wgamma_prl}, so that we
expect about 0.018 events due to the extension of the selection cuts. 
We also have to take into account that we specifically extended the selection
criteria to add this event. The additional background only contributes to
experiments in which at least one event satisfies the extended
selection cuts. This is expected to happen only once every six experiments. The
additional background component is therefore six times 0.018 or 0.11 events. 
The most significant source of these background events are \ttbar\ decays to
$e$+jets with a muon-tagged jet, in which one jet is misidentified as an
electron.

In total, two $ee$ events enter our final sample.

\subsection {$\mu\mu$ Channel}

The dimuon channel shares the $Z \rightarrow \ell\ell$ background with the
dielectron channel. The less precise measurement of the muon momentum makes
separation of the \ttbar\ signal from this background more difficult. In order
to reduce this background,  a kinematic fit to the  $Z\to\mu\mu$ hypothesis  is
applied, and the event is required to have $\chi^2$ probability less than 1\%
for this fit.
Even after this cut, $Z$ boson production
remains the dominant background source. Instrumental backgrounds arise from
heavy quark jets with a high-\pt\ muon that is misidentified as an isolated
muon.

One event survives all selection criteria.

\subsection {Dilepton Events}

Six events enter our dilepton event sample: three are $e\mu$
events, two are $ee$ events, and one is a $\mu\mu$ event. Table 
\ref{tab:em1} lists the properties of these events. 

\begin{table}
\begin{center}
\caption{Kinematic properties of dilepton events (momenta in \gevc)
used in the reconstruction of the top quark mass. All corrections
are included.
\label{tab:em1}}
\begin {tabular}{llrrrrrr}
Event & Object & $p_x$ & $p_y$ & $p_z$ & $p_T$ & $\eta$ & $\phi$ \\ 
\hline
$e\mu$\#1 & $e$    & $ 12.3$ & $-97.8$ & $ 41.1$ & $ 98.6$ & $  0.41$ & $  4.84 $\\
          & $\mu$  & $-68.3$ & $272.5$ & $ 95.1$ & $280.0$ & $  0.33$ & $  1.82 $\\ 
          & \mpt   & $100.5$ &$-152.7$ &   ---   & $182.9$ &    ---   & $  5.29 $\\ 
          & jet    & $-25.5$ & $ -9.9$ & $-20.8$ & $ 27.3$ & $ -0.70$ & $  3.51 $\\
          & jet    & $-14.4$ & $-20.5$ & $ 32.3$ & $ 25.1$ & $  1.07$ & $  4.10 $\\
\hline                                                                            
$e\mu$\#2 & $e$    & $-75.4$ & $ -1.1$ & $-30.2$ & $ 74.5$ & $-0.39 $ & $ 3.16  $\\
          & $\mu$  & $-25.2$ & $ 10.6$ & $-12.8$ & $ 27.4$ & $-0.45 $ & $ 2.75  $\\
          & \mpt   & $ 62.0$ & $  5.2$ &   ---   & $ 62.3$ &   ---    & $ 0.08  $\\ 
          & jet    & $ 38.9$ & $-85.6$ & $-16.0$ & $ 94.0$ & $-0.17 $ & $ 5.14  $\\
          & jet    & $ 14.2$ & $ 33.1$ & $-11.4$ & $ 36.0$ & $-0.31 $ & $ 1.17  $\\
          & jet    & $ -1.6$ & $ 29.3$ & $ 11.9$ & $ 29.4$ & $ 0.39 $ & $ 1.63  $\\
\hline                                                                            
$e\mu$\#3 & $e$    & $-44.7$ & $ 20.2$ & $140.1$ & $ 49.1$ & $  1.77$ & $  2.72 $\\
          & $\mu$  & $  5.4$ & $ 17.2$ & $ -3.3$ & $ 18.1$ & $ -0.18$ & $  1.27 $\\
          & \mpt   & $-12.5$ & $  4.5$ &   ---   & $ 13.2$ &   ---    & $  2.79 $\\
          & jet    & $ 39.6$ & $-29.9$ & $ 11.3$ & $ 49.7$ & $  0.22$ & $  5.64 $\\
          & jet    & $ 19.8$ & $-19.4$ & $-31.0$ & $ 27.7$ & $ -0.97$ & $  5.51 $\\
\hline                                                                            
$ee$\#1   & $e$    & $  2.7$ & $ 50.4$ & $ 17.1$ & $ 50.5$ & $  0.33$ & $  1.52 $\\
          & $e$    & $ -7.4$ & $ 21.4$ & $-47.6$ & $ 22.6$ & $ -1.49$ & $  1.91 $\\
          & \mpt   & $ 41.3$ & $ -4.0$ &   ---   & $ 41.5$ &   ---    & $  6.19 $\\
          & jet    & $-29.2$ & $-36.9$ & $-37.0$ & $ 47.1$ & $ -0.72$ & $  4.04 $\\
          & jet    & $  3.5$ & $-27.1$ & $-28.9$ & $ 27.4$ & $ -0.92$ & $  4.84 $\\
\hline                                                                            
$ee$\#2   & $e$    & $ 52.3$ & $ -4.1$ & $-34.4$ & $ 52.5$ & $ -0.62$ & $  6.20 $\\
          & $e$    & $ -8.5$ & $-26.6$ & $ 27.0$ & $ 27.9$ & $  0.86$ & $  4.40 $\\ 
          & \mpt   & $ 42.6$ & $-11.3$ &   ---   & $ 44.1$ &   ---    & $  6.02 $\\
         & jet$^*$ & $-92.4$ & $-26.0$ & $-61.6$ & $ 96.0$ & $ -0.60$ & $  3.41 $\\
          & jet    & $-23.5$ & $ 25.3$ & $-34.0$ & $ 34.6$ & $ -0.87$ & $  2.32 $\\
          & jet    & $  0.0$ & $ 27.7$ & $ 18.3$ & $ 27.7$ & $  0.62$ & $  1.57 $\\
\hline                                                                            
$\mu\mu$  & $\mu$  & $-63.9$ & $ 12.7$ & $-21.4$ & $ 65.1$ & $ -0.32$ & $  2.94 $\\
          & $\mu$  & $-16.0$ & $ 31.0$ & $  1.9$ & $ 34.9$ & $  0.05$ & $  2.05 $\\
          & \mpt   & $ 71.2$ & $ 53.2$ &   ---   & $ 88.9$ &   ---    & $  0.64 $\\
          & jet    & $ 33.8$ &$-103.1$ &$-107.6$ & $108.5$ & $ -0.88$ & $  5.03 $\\
          & jet    & $ -9.1$ & $ 22.7$ & $ 27.7$ & $ 24.5$ & $  0.97$ & $  1.95 $\\
          & jet    & $ -8.4$ & $-18.6$ & $ 47.8$ & $ 20.5$ & $  1.58$ & $  4.29 $\\ 
\end {tabular}
$^*$ tagged by a soft muon 
\end{center}
\end{table}

\section{ Reconstruction of the Top Quark Mass}
\label{sec-mt}
\subsection {Characteristics of Dilepton Events}

\begin{figure}[htpb]
\centerline{\psfig{figure=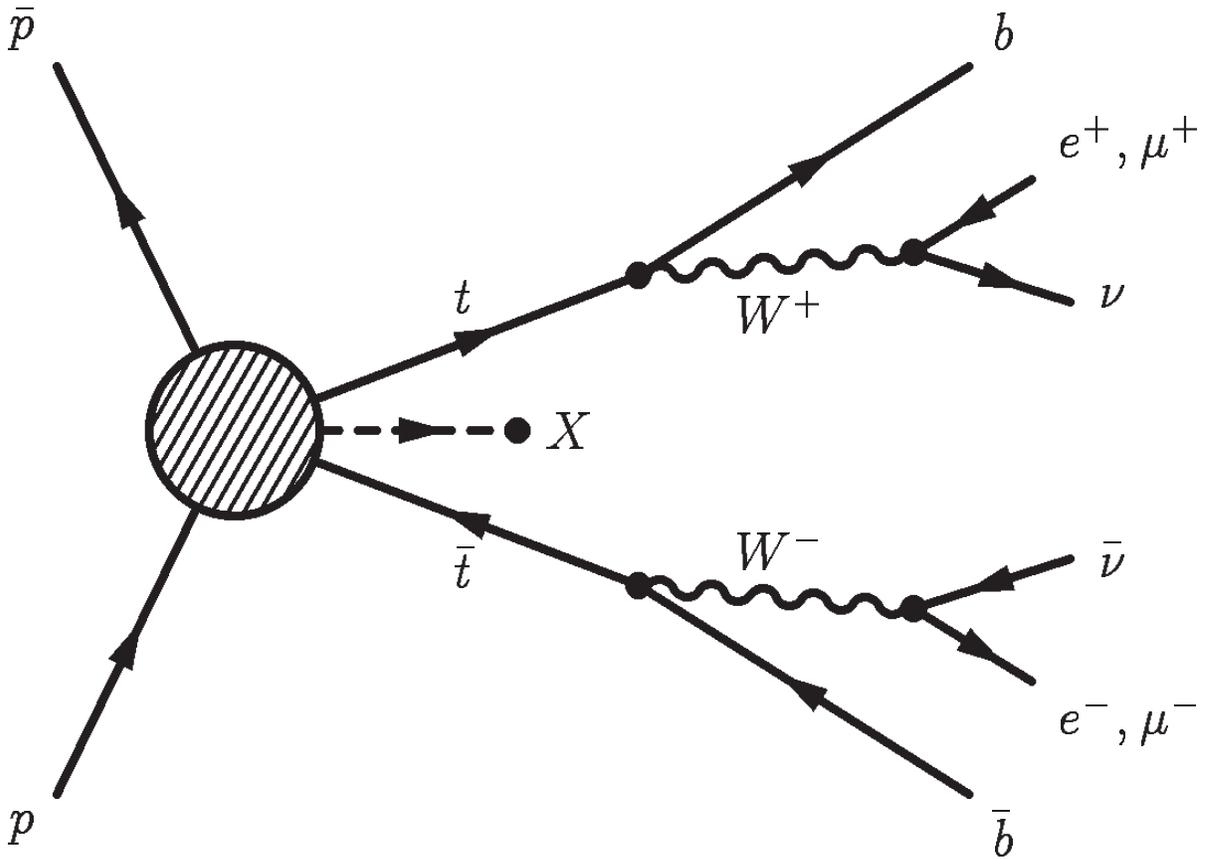,width=\hsize}}
\caption {Schematic representation of \ttbar\ production and decay in the
dilepton channels.}
\label{fig:dilep_cartoon}
\end{figure}

The dilepton decay topology does not provide sufficient
information to uniquely reconstruct the $t$ and $\tbar$ quarks.  In the
simplest scenario, the decay $t\to W^+b$, $\tbar\to W^-\bbar$, followed by
$W^+\to\l^+\nu$ and $W^-\to\l^-\nubar$ produces six particles in the final
state: two charged leptons, which we allow to be either
electrons or muons ($ee$, $e\mu$, or $\mu\mu$); two neutrinos ($\nu$,$\nubar$);
and two $b$ quarks ($b$,\bbar), as shown in Fig.~\ref {fig:dilep_cartoon}.
Given the identities of the particles, this final state is therefore completely
specified by the momenta of these six particles, \ie\ 18 numbers. We measure
the momenta of the charged leptons and the jets from the hadronization
of the $b$ quarks directly. In addition, the observed \mptv\ provides
the $x$ and $y$
components of the sum of the neutrino momenta for a total of 14 measurements.
Assuming $m_t > \mW+m_b$ we can
impose three constraints, two on the masses of the decaying $W$ bosons,
$m^{\l^+\nu} = m^{\l^-\nubar} = \mW$, and one on the masses of the top quarks,
$m^{\l^+\nu b} = m^{\l^-\nubar\bbar}$. This leaves us with 17 equations and 18
unknowns so that a kinematic fit would be underconstrained. We have to
develop a different procedure to obtain an estimate of the top quark mass from
the available information. This is the fundamental difference between the
mass determination in the dilepton channel and that in the lepton+jets channel,
which allows a kinematic fit with two constraints.

We solve this problem by fitting the dynamics of the decays \cite{Kondo}. For
each event we derive a weight function, which is a measure of the probability
density for a \ttbar\ pair to decay to the observed final state, as a function
of the top quark mass. We compare these weight functions to Monte Carlo
simulations of \ttbar\ decays for different values of the top quark mass and use
a maximum likelihood fit to extract the mass value that yields the best
agreement.

\subsection {Computation of the Weight Function}

Ideally we would like to compute analytically the
probability density for a \ttbar\ pair to decay to the observed final state
for any given value of the top quark mass. This probability density is given by
\begin{eqnarray}
\P(\o|m_t) \propto 
\int f(x)f(\xbar)|\M|^2 p(\o|\v)\delta^4\d^{18}\v\,\d x\, \d\xbar,
\label{eq:dilep_exact_prob}
\end {eqnarray}
where \o\ is the set of 14 measured quantities and \v\ is the set of 18
parameters
that specify the final state. \M\ is the matrix element for the process $\qqbar$
or $gg\to\ttbar+X\to\l^+\nu b\l^-\nubar\bbar+X$, $f(x)$ the parton density for
quarks or gluons of momentum fraction $x$ in the proton, and $f(\xbar)$ that for
antiquarks or gluons of momentum fraction $\xbar$ in the antiproton. The
detector resolution function $p(\o|\v)$ is the probability density to observe
the values \o\ given the final state parameters \v. The four-dimensional
$\delta$-function enforces the four mass constraints:
\begin{eqnarray}
\delta^4=\delta(m^{\l^+\nu}-\mW) \times\delta(m^{\l^-\nubar}-\mW) \times
         \delta(m^{\l^+\nu b}-m_t) \times\delta(m^{\l^-\nubar\bbar}-m_t).
\end {eqnarray}
Here we neglect the finite widths of the $W$ boson and the top quark.

Unfortunately this expression involves a multidimensional integral that has
to be evaluated numerically and is complicated by the need to include initial
and final state gluon radiation. Such higher order effects complicate the
reconstruction of the top quark mass substantially and cannot be neglected.
We therefore do not attempt to compute the exact probability density given in
Eq. \ref{eq:dilep_exact_prob}. Rather, we construct simpler weights that
retain sensitivity to the value of the top quark mass but can be evaluated with
the available computing resources. We calibrate the effect of the
simplifications by comparing the weight functions obtained from the collider
data to Monte Carlo simulations (Sect. \ref{sec-fit}).

The calculation of the weight function proceeds in three steps.
First we map the observed charged  leptons and jets to the corresponding $t$ and
$\tbar$   decay  products.  There  are   ambiguities in  this  step  because the
fragmentation of the $b$ quarks may result in more than one reconstructed jet or
because a  gluon radiated  from the  initial state  may  contribute a jet to the
event. We  cannot, in  general,  distinguish between jets  originating  from
gluons and
quarks. Furthermore, we do not measure the sign of the electron charge nor can we
distinguish between  jets originating  from quarks and  antiquarks. Therefore,
there is an ambiguity in pairing the charged leptons and $b$ jets that originate
from the  same top  quark. We  repeat the  following two  steps for  each of the
possible assignments and add the resulting weight functions.

Given the  charged lepton and  $b$ quark  momenta from the  decay of the $t$ and
$\tbar$     quarks  and  the  sum  of  the    neutrino   momentum    components,
$\px^{\nu\nubar}$  and  $\py^{\nu\nubar}$, we compute a  weight as a function of
the top  quark mass.  We have  developed two  algorithms  to  compute the weight
function which  emphasize different  aspects of  top production  dynamics. The
first  algorithm   (matrix-element   weighting) is  an  extension of  the weight
proposed in Ref. \cite{DG} and takes into  account the parton distribution
functions for the  initial proton and  antiproton and the  decay distribution of
the $W$ bosons due to the  $V$--$A$ coupling of the  charged current. The second
(neutrino weighting)  \cite{Varnes_thesis} is based on the available phase space
for neutrinos from the decay of the \ttbar\ pair.

Finally we average the weight function over the experimental resolution.

In the following, we first discuss
the ambiguities in associating the observables with final state particles.
Then we discuss the two algorithms that are used to compute
the weight functions and finally the experimental resolutions.

\subsection {Jet Combinatorics}
\label {sec-jet_comb}

In the calorimeter we detect the jets from the fragmentation of the two $b$
quarks. The fragmentation of a $b$ quark can produce more than one jet because
of hard gluon radiation. This corresponds to final state radiation. Jets can
also originate from gluons radiated by partons in the initial state. We refer to
this as initial state radiation. It is not possible to tell whether a jet
originates from the fragmentation of a quark or a gluon, unless a $b$ quark
decays semileptonically to a muon that we subsequently detect. Thus,
reconstruction of the original partons from the observed jets presents some
complication.

We consider jets with $p_T>15$ \gevc. If there are only two such jets we
assign their measured momenta to the two $b$ quarks. If there are more than two
jets we have a range of possible assignments. To limit the possibilities, we
restrict the procedure to the three leading jets in $p_T$. We assign two of them
to the $b$ quarks and the third jet either to initial state radiation, in which
case we ignore it, or to final state radiation, in which case we add its
momentum to that of one of the two $b$ quarks. There are six possible
permutations for three jets, as listed in Table \ref{tab:dilep_jet_perms}.

\begin{table}
\begin{center}
\caption {Possible assignments of three observed jets ($j_1$, $j_2$, and
$j_3$) to the $b$ quarks and initial state radiation (ISR).
\label{tab:dilep_jet_perms}}
\begin {tabular}{cccc}
Permutation & \multicolumn{2}{c}{$b$-Jets} & ISR   \\ \hline
1	    & $j_1$     & $j_2$     & $j_3$ \\
2	    & $j_1$     & $j_3$     & $j_2$ \\
3           & $j_2$     & $j_3$     & $j_1$ \\
4           & $j_1+j_2$ & $j_3$     & ---   \\
5           & $j_2+j_3$ & $j_1$     & ---   \\
6           & $j_1+j_3$ & $j_2$     & ---   \\
\end {tabular}
\end {center}
\end {table}

If there is a jet in the event that is tagged by a soft muon,
we only allow permutations that assign this jet to a $b$ quark.
In the collider data sample this is the case for one $ee$ event.

Not all permutations are equally likely to be correct. For each jet considered
to be due to initial state radiation, we assign a weight factor
\begin{equation}
\label{eq:Qisr}
\Q_{\rm ISR} = \exp\left({-p_T^{j}\sin\theta^{j}\over25\ \gevc}\right).
\end{equation}
Similarly, for every pair of jets that is assigned to a $b$ quark, we define
\begin{equation}
\label{eq:Qfsr}
\Q_{\rm FSR} = \exp\left({-m^{jj}\over20\ \gevcc}\right),
\end {equation}
where $m^{jj}$ is the invariant mass of the two jets. These functional forms of
the weights were derived empirically from a study of \ttbar\ decays generated
by \ISAJET\ \cite{isajet}. The factor $\Q_{\rm ISR}$ favors assignments in  which
jets from initial state radiation are close to the beam direction, and
$\Q_{\rm FSR}$ favors the merging of jets which are soft or close together. The
numerical coefficients of the exponents are chosen such that the mean
reconstructed top quark masses for events with two-jet and multi-jet final
states are the same.

After adding the four-momenta of the jets assigned to a $b$ quark, we rescale
the momentum components, keeping the energy fixed, so that the $b$ quark 
four-momentum has an invariant
mass of 5 \gevcc\ to put the outgoing quark momentum on the mass shell.

There are two ways to pair the momenta of the two charged leptons with the two
$b$ quark momenta. Since we cannot determine which $b$ quark originated from the
decay of the $t$ quark and which from the decay of the $\tbar$ quark, we
consider both pairings with equal probability.

\subsection {Matrix-Element Weighting (\MWT) Algorithm}

Assuming that we know the momenta of the charged leptons ($p^{\l^+}$,
$p^{\l^-}$), the $b$ quarks ($p^b$, $p^{\bbar}$), and the sum of the $x$ and $y$
components of the neutrino momenta ($\px^{\nu\nubar}$, $\py^{\nu\nubar}$) and
that we impose the three constraints mentioned above, we are still one constraint
short of being able to solve for the unknown components of the neutrino
momenta. Assuming a fixed value for the top quark mass $m_t$ supplies the
required constraint to solve the problem, except for a fourfold ambiguity.
Not all solutions are equally likely for any given value of $m_t$. We therefore
assign a weight to the $i^{th}$ solution \cite{DG}:
\begin {equation}
w^\M_i(m_t) = f(x)f(\xbar)\,p(E^{\l^-*}_i|m_t)\, p(E^{\l^+*}_i|m_t),
\end {equation}
where $f(x)$ and $f(\xbar)$, the parton distribution functions, are evaluated at
$Q^2=m_t^2$, and $p(E^{\l*}|m_t)$ is the probability density function for the
energy of the charged lepton in the rest frame of the top quark ($E^{\l*}$).
This probability density is given by
\begin {eqnarray}
p(E^{\l*}|m_t) = 
 { 4 m_t E^{\l*} ( m_t^2 - m_b^2 - 2 m_t E^{\l*})
\over ( m_t^2 - m_b^2 )^2 + \mW^2 ( m_t^2 + m_b^2 ) - 2 \mW^4 }.
\end{eqnarray}

We sum the weights for all solutions and normalize by a factor ${\cal A}(m_t)$
to obtain the weight for the event
\begin{equation}
w^\M(m_t) = {\cal A}(m_t) \sum_{i=1}^4 w^\M_i(m_t).
\end{equation}
The factor ${\cal A}(m_t)$ ensures that the average weight is independent 
of the top quark mass.
We compute the weight function for $82\lt m_t\lt278$ \gevcc\ in steps of 4
\gevcc, where the lower limit is given by the requirement that the top quark
decays into a real $W$ boson and a $b$ quark and the upper limit is placed well
above the measurement of the top quark mass in the lepton+jets channel.
The normalization factor is computed using a Monte Carlo simulation so that
\begin{equation}
\sum_{1}^N w^\M(m_t) = N,
\end {equation}
where the sum is over the events that pass the
selection cuts. We parametrize the  factor ${\cal A}(m_t)$ 
at different values of $m_t$ (in \gevcc) as
\begin{equation}
{\cal A}(m_t) = \left(5.86 - 0.044 m_t + 0.000084 m_t^2 \right)^{-1}.
\end{equation}

\subsection{Neutrino Weighting (\vWT) Algorithm}

The neutrino weighting
algorithm also computes a weight as a function of the top quark mass.
In contrast to the \MWT\ algorithm it does not solve for the unknown neutrino
momentum components, but rather samples the neutrino pseudorapidity space and computes
a weight based on how much of the sampled space is consistent with the observed
\mpt.

For every value of the top quark mass, we sample the rapidities of neutrino
($\eta^\nu$) and
antineutrino ($\eta^{\nubar}$) from the \ttbar\ decay. For each top
decay we then know the momenta of the charged lepton and the $b$ quark, the
assumed neutrino pseudorapidity, and the top quark mass, which allows us to solve for
the
transverse momentum components of the neutrino ($p_x^\nu$ and $p_y^\nu$) with
a twofold ambiguity. The two solutions for each of the two top decays combine
to give four solutions for the event. For the $i^{th}$ solution we compute a
weight
based on the agreement between the observed \mpt\ and the sum of the calculated
neutrino \pt\ values:
\begin{eqnarray}
w^\nu_i(m_t) &=& \exp{\left(-\left(\mpx-p_x^\nu-p_x^{\nubar}\right)
^2\over2\sigma^2\right)\times} 
\exp{\left(-\left(\mpy-p_y^\nu-p_y^{\nubar}\right)^2\over2\sigma^2\right)},
\end {eqnarray}
where $\sigma=4$ \gevc\ is the resolution for each component of \mptv\ (Sect.
\ref{sec-mt_res}).

\begin{figure}[htpb]
\centerline{\psfig{figure=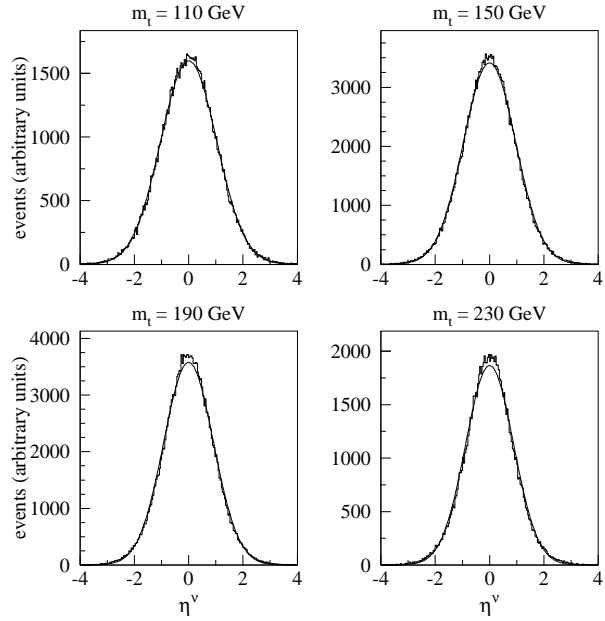,height=3.5 in}}
\caption {Distributions of neutrino pseudorapidity from top quark decay,
modeled by \HERWIG, for several top quark masses. The smooth curves are fits to
Gaussians.
\label{fig:hw_nu_eta}}
\end{figure}

\begin{figure}[htpb]
\centerline{\psfig{figure=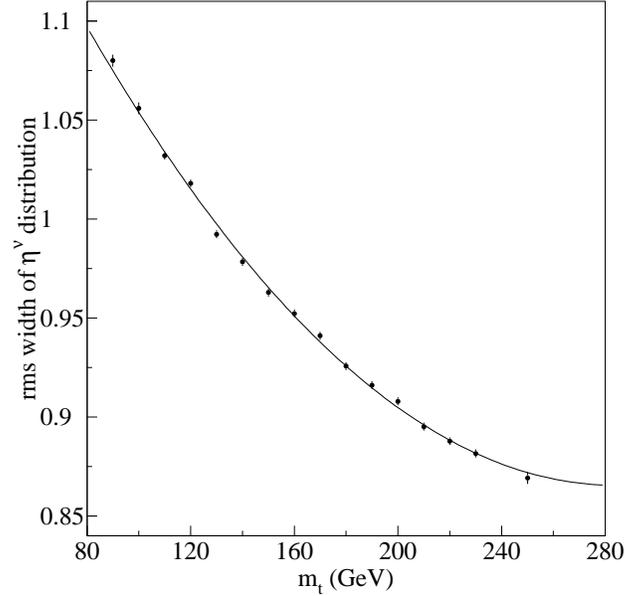,height=3.5 in}}
\caption {  Width of the Gaussian curves fit to the neutrino pseudorapidity
distributions as a function of top quark mass.  The smooth line is the
polynomial parametrization used in the analysis.
\label{fig:hw_nu_eta_fit}}
\end{figure}

Not every value of the neutrino pseudorapidity is equally likely. Figure
\ref{fig:hw_nu_eta} shows the distribution of neutrino rapidities predicted by
the \HERWIG\ Monte Carlo program for several top quark
masses. The distributions can be approximated by Gaussian curves. The width
$\sigma_\eta$ of the Gaussian varies as a function of the top quark mass. It can
be parametrized by the second order polynomial 
\begin{equation}
\sigma_\eta = 5.56\times10^{-6}m_t^2 -2.16\times10^{-3}m_t +1.314,
\end{equation}
as shown in Fig. \ref {fig:hw_nu_eta_fit}. We compute the weights $w^\nu_i$ for
ten values of each of the neutrino rapidities, spaced such that they divide the
Gaussian into slices of equal area.

To obtain the weight for the event we add the weights for all four solutions
and all values of the neutrino rapidities,
\begin{equation}
w^\nu(m_t) = \sum_{\eta^\nu}\sum_{\eta^{\nubar}}\sum_{i=1}^4 w^\nu_i(m_t).
\end{equation}

\subsection {Detector Resolution}
\label{sec-mt_res}

The algorithms described in the two previous sections use as input the
measured momenta of the charged leptons and $b$ jets and the transverse
components of the sum of the neutrino momenta. To account for finite
resolution, we integrate the weights over the ranges of these
quantities that are consistent with the measurements to smooth out the
weight functions.

To evaluate this integral, we generate a large number of sets of event
parameters over which we average the weights. These sets of event parameters
derive from the observed events by adding normally distributed resolution terms
to the observed values to populate the parameter space consistent with
the measured values. The new values $\tilde o$ are given in terms of the
observed value $o$, the resolution, $\sigma$, for the measurement of $o$, and a
normally distributed random variable $\xi$:
\begin {equation}
\tilde o = o + \sigma \xi.
\end {equation}
We apply such fluctuations to all momentum measurements. Directions
are relatively precise and are therefore not fluctuated. This also reduces the
number of numerical operations.

The energy resolution for electrons is
\begin{equation}
\sigma(E^e) = 0.15\ \gevc^{1\over2}\sqrt{E^e}.
\label{eq:em_res}
\end{equation}
The resolution function for the inverse of the muon momentum is approximately
Gaussian. We therefore fluctuate the inverse of the momentum with the
resolution
\begin{equation}
\sigma\left({1\over p^\mu}\right) = \left\{\left(0.18 (p^\mu-2\ \gevc)\over
{p^\mu}^2\right)^2 + \left({0.003\over\gevc}\right)^2 \right\}^{1\over2}.
\label{eq:mu_res}
\end{equation}

The energy resolution for jets receives contributions from several effects.
One is the intrinsic resolution of the calorimeter. The energy of the jet is
measured as the energy in a cone of radius $\Delta R=0.5$. This energy is not
identical to that of the parton. Additional energy can be accrued from
overlap with other jets and energy can be lost due to gluon radiation outside of
the cone. These contributions to the resolution depend on
the process and we therefore use Monte Carlo \ttbar\ events
to evaluate the jet energy resolution.

\begin{figure}[htpb]
\centerline{\psfig{figure=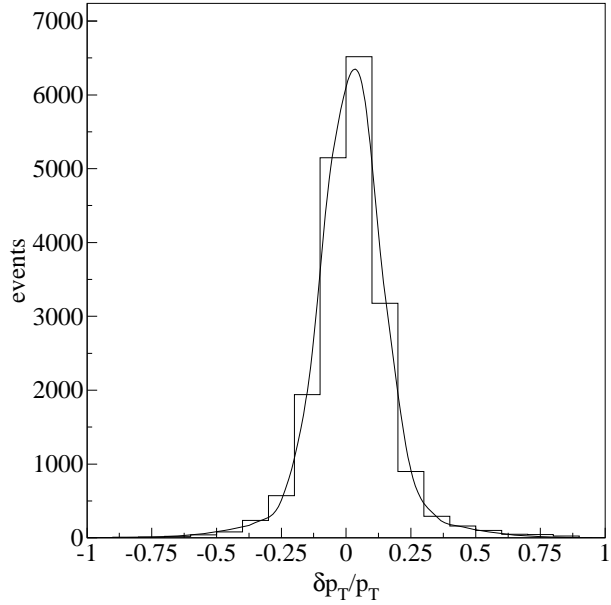,height=3.5 in}}
\caption {Fractional \pt\ resolution for jets with $50\lt
\pt\lt60$ GeV from
\ttbar\ decays generated with top quark masses between 110 and 190
\gevcc\ using the \HERWIG\ program. The superimposed curve is the fit using two
Gaussian curves.
\label{fig:jet_res_parm}}
\end{figure}

\begin{figure}[htpb]
\centerline{\psfig{figure=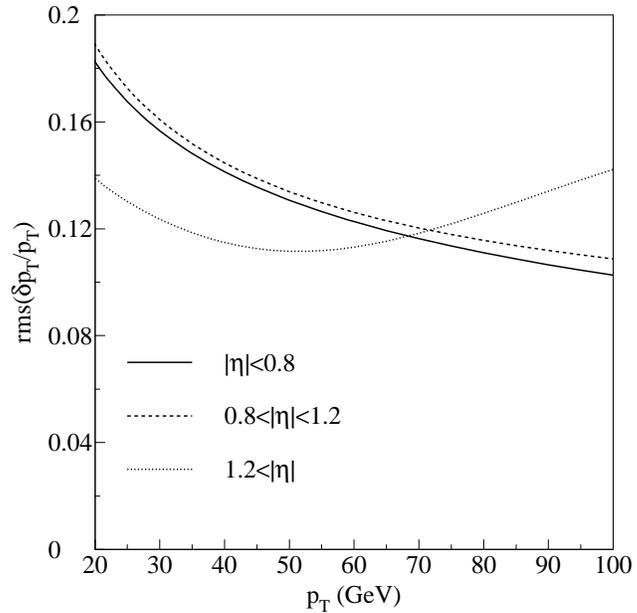,height=3.5 in}}
\caption {Rms width of fractional jet \pt\ resolution functions versus jet
\pt\ for three pseudorapidity regions.
\label{fig:jres}}
\end{figure}

We compare the reconstructed jet \pt\ to that of the nearest cluster of
hadrons generated by the Monte Carlo in a sample of \ttbar\ events with top
quark masses ranging from 110 to 190 \gevcc. Typically, the distribution in the
fractional mismeasurement in \pt\ exhibits a narrow peak due to
the intrinsic calorimeter resolution and broad tails due to ambiguity in the
jet definition.  We fit two Gaussian curves with equal means but different
widths to the distribution, and parametrize the widths of the two Gaussians and
their relative normalization as functions of \pt\ and $\eta$. Figure
\ref{fig:jet_res_parm} shows a typical distribution along with the fit that we
use as a resolution function. Figure \ref{fig:jres} shows the rms resolution as
a function of \pt.

The Monte Carlo simulation used to determine the jet energy resolution neither
includes noise due to the intrinsic radioactivity of the uranium nor due to
multiple interactions. We therefore add an additional
uncorrelated constant noise term of 5--6 \gev, depending on $\eta$. These
values were determined by balancing the \pt\ vectors in dijet events.

Using a sample of random \ppbar\ interactions, we measure the resolution for
any component of \mptv\ to be about 4 \gevc. Both components of \mptv\ are
fluctuated by this resolution. The \mpt\ vector is also corrected for the
fluctuations in the lepton and jet momenta.

The number of variations performed for each event is limited by the available
computing power. We average over 100 variations per event for Monte
Carlo samples and 5000 variations per event for the collider data.

The weight function for each event is then
\begin{equation}
W^x(m_t)={1\over N'N''}\sum_{j=1}^{N'}\sum_{k=1}^2
\sum_{l=1}^{N''}\Q_{\rm ISR}\Q_{\rm FSR}w^x(m_t),
\end{equation}
where $\Q_{\rm ISR}$ and $\Q_{\rm FSR}$ are the parametrized weights defined in
Eqs. \ref{eq:Qisr} and \ref{eq:Qfsr}. The index $j$ runs over the $N'$
resolution fluctuations, $k$ over the two lepton--$b$ jet pairings, $l$ over
the $N''$ jet permutations, and $x$ refers to the \MWT\ or \vWT\ algorithms.

\begin{figure}[htpb]
\centerline{\psfig{figure=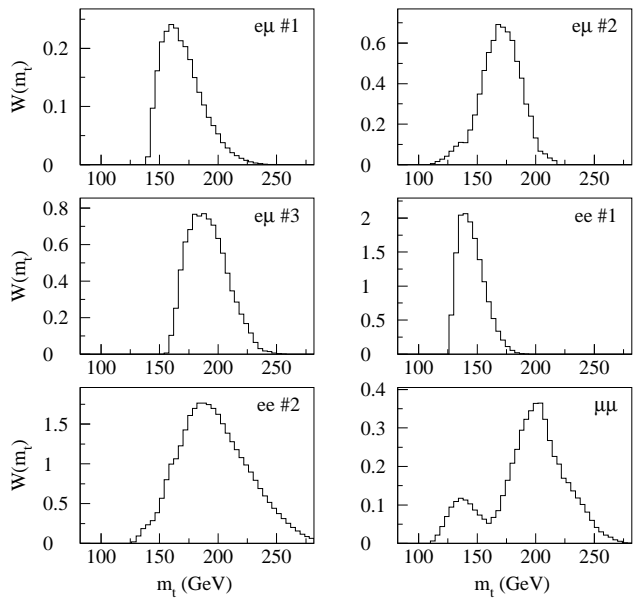,height=3.5 in}}
\caption {$W(m_t)$ functions for the dilepton events from the \MWT\ analysis.
The labels in the upper right hand corners identify the events (cf. Table
\ref{tab:em1}).
\label{fig:dilep_cand_wts_m}}
\end{figure}

\begin{figure}[htpb]
\centerline{\psfig{figure=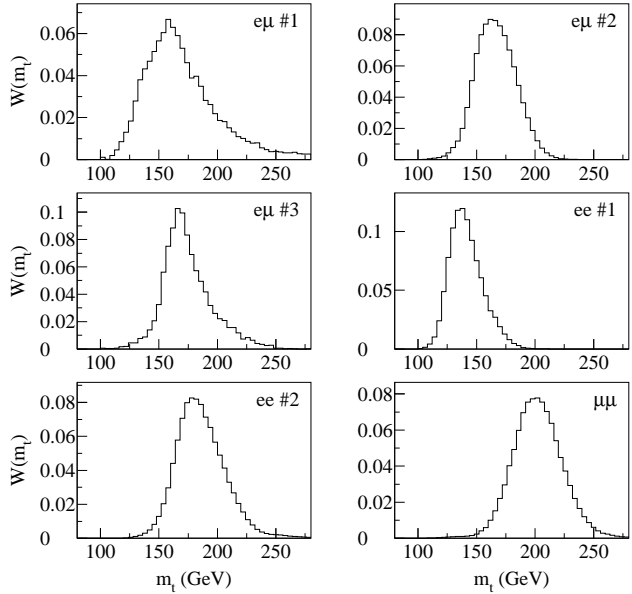,height=3.5 in}}
\caption {$W(m_t)$ functions for the dilepton events from the \vWT\
analysis. The labels in the upper right hand corners identify the events (cf.
Table \ref{tab:em1}).
\label{fig:dilep_cand_wts}}
\end{figure}

Figure \ref{fig:dilep_cand_wts_m} shows $W(m_t)$ for the dilepton events for
the \MWT\ analysis and Fig.~\ref{fig:dilep_cand_wts} shows the corresponding
functions for the \vWT\ analysis.

\subsection {Monte Carlo Tests}

We now describe tests of the properties of the weight functions
to demonstrate their sensitivity to the top quark mass and other parameters.

\subsubsection{Parton-level Tests}

Parton-level tests are based on the momenta of the partons
generated by the Monte Carlo simulation.
Tests at this level are neither subject to effects from
detector resolution nor initial or final state radiation.
To restrict the sample to events that are broadly similar to those which enter
the collider data analysis, the event selection for these tests requires
two $b$ quarks and  two leptons with $p_T > 20$ \gevc\ and $|\eta| < 2.5$.

\begin{figure}[htpb]
\centerline{\rlap{\raisebox{2.9in}{\hspace{0.6in}(a)}}
            \rlap{\raisebox{1.3in}{\hspace{0.6in}(b)}}
            \psfig{figure=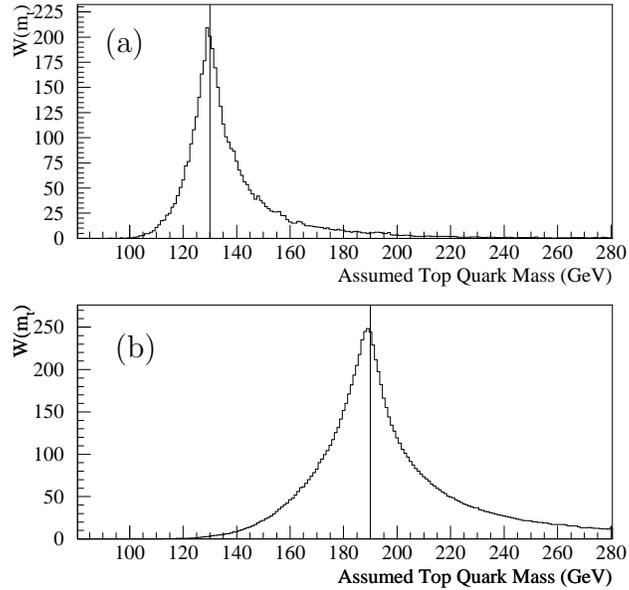,height=3.5 in}}
\caption {Average parton-level weight $W(m_{t})$ for \ttbar\ decays with (a)
$m_t=130$ \gevcc\ and (b) $m_t=190$ \gevcc\ for the \vWT\ algorithm. The
vertical lines indicate the input mass values.
\label{fig:parton_level_wt}}
\end{figure}

We examine the average weight function  as a function of input top quark mass by
normalizing the  area of the  weight function  for each event  to unity and then
summing these  normalized  functions for a  collection of Monte  Carlo events. A
sample of  10{,}000 events was  used, about  half of which  passed the cuts. The
results are shown in Fig.~\ref{fig:parton_level_wt} for top quark masses of 130
and 190  \gevcc. On  average, the  weight  function is  sharply  peaked within
one \gevcc\ of the  input mass. The tails  of the function are  asymmetric, with
the high-end tail extending further than the low-end tail.

\begin{figure}[htpb]
\centerline{\psfig{figure=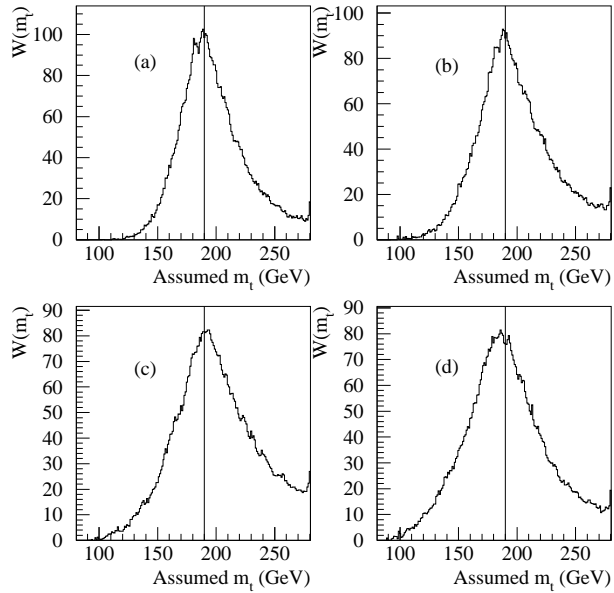,height=3.5 in}}
\caption {Average parton-level weight functions for the \vWT\ algorithm,
obtained (a) with the parton momenta smeared by the detector resolutions, (b)
with the two-fold ambiguity in lepton-jet pairings included, (c) with ISR but
without FSR, and (d) without ISR but with FSR. The vertical lines indicate
the input mass value of 190 \gevcc.
\label{fig:dilep_radiation}}
\end{figure}

Figure \ref{fig:dilep_radiation} shows the impact of detector resolution, jet
combinatorics, and radiation on the weight functions for 190 \gevcc\ Monte
Carlo events. The distribution becomes significantly broader when resolution
effects and both lepton-$b$ jet pairings are considered, but
the peak value remains unchanged. Initial state radiation
increases the mean value and adds a high-mass tail, as expected. Final state
radiation has the opposite effect.  In total, the effect of resolution,
combinatorics, and radiation is to broaden the distribution of the weight
function and move the peak of the distribution away from the input mass.

\subsubsection{Tests using Full Simulation}

To quantitatively assess the response of the fitting algorithm to
events from the D\O\ data sample that pass the kinematic selection described in
Sect. \ref {sec-evt}, we use fully simulated samples of \HERWIG\
\ttbar\ decays. In contrast to the parametrized detector response used in the
parton-level tests, these samples derive from a detailed detector model
implemented using the \GEANT\ program. The events are processed with the same
reconstruction program and filtered using the same kinematic criteria as for the
collider data.

\begin{figure}[htpb]
\centerline{\psfig{figure=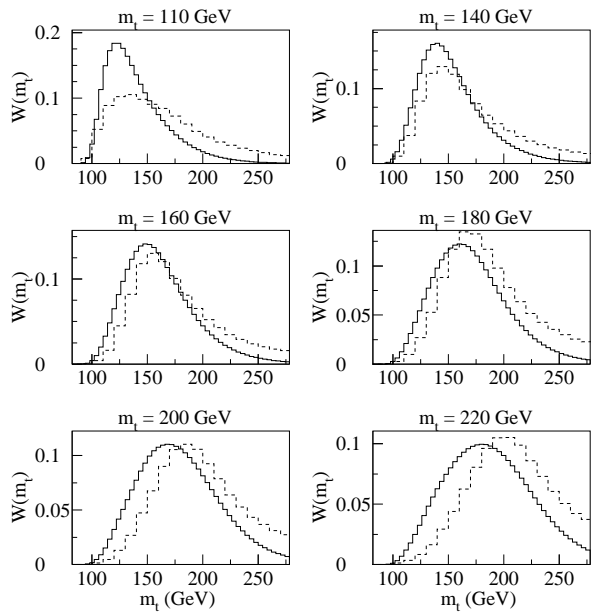,height=3.5 in}}
\caption{Average weight functions for fully simulated \ttbar\ decays
events in the $e\mu$ channel from the \MWT\ analysis (solid line) and
the \vWT\ analysis (dashed line).
\label{fig:emu_fits}}
\end{figure}

\begin{figure}[htpb]
\centerline{\psfig{figure=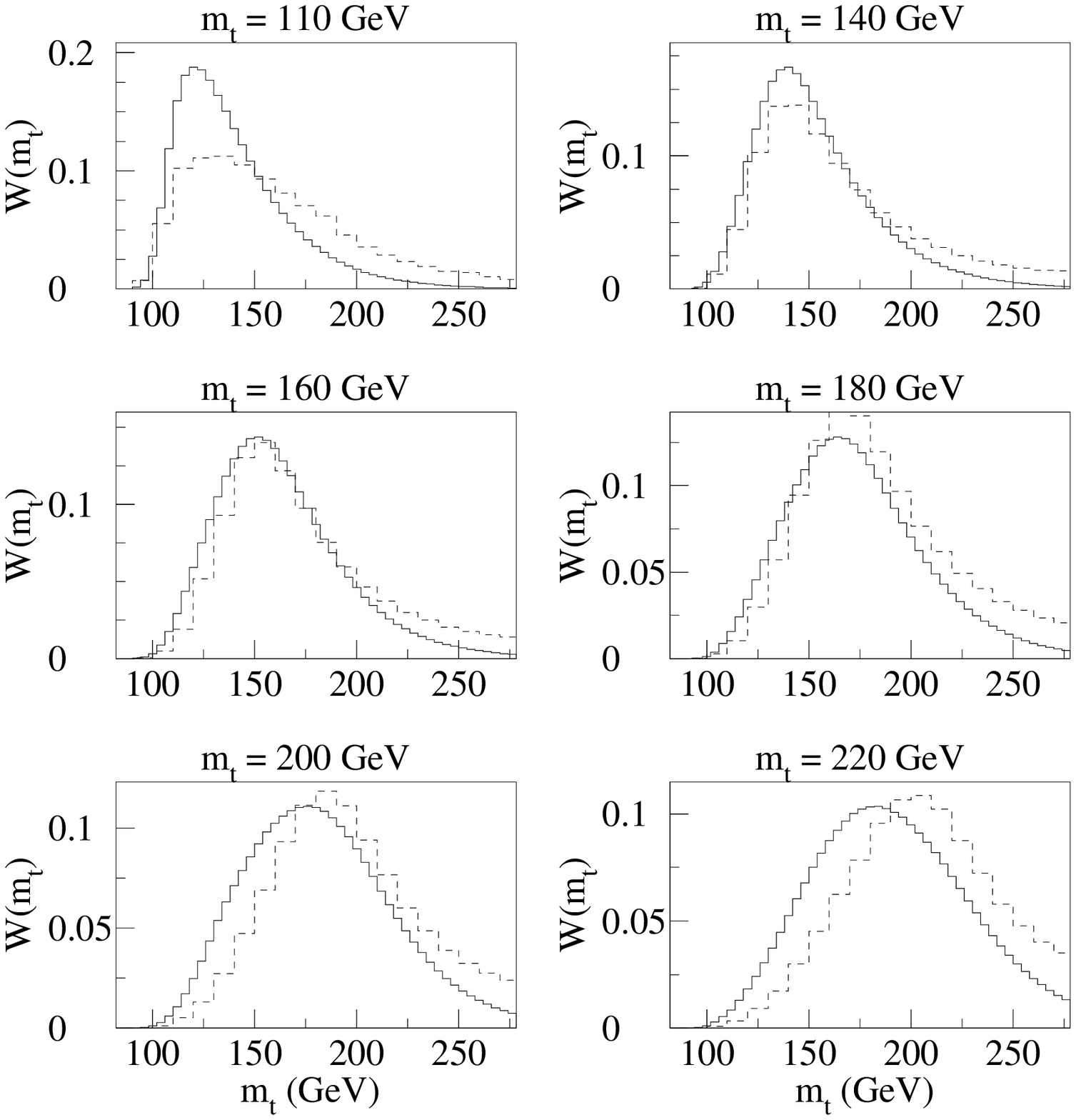,height=3.5 in}}
\caption{Average weight functions for fully simulated \ttbar\ decays
in the $ee$ channel from the \MWT\ analysis (solid line) and
the \vWT\ analysis (dashed line).
\label{fig:ee_fits}}
\end{figure}

\begin{figure}[htpb]
\centerline{\psfig{figure=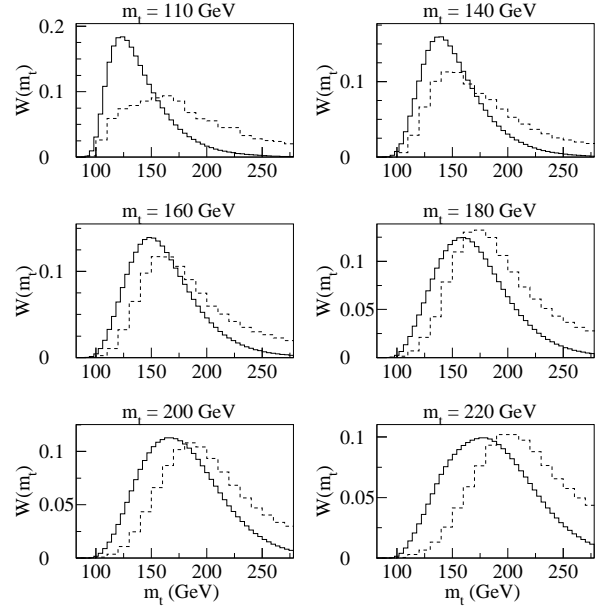,height=3.5 in}}
\caption{Average weight functions for fully simulated \ttbar\ decays
events in the $\mu\mu$ channel from the \MWT\ analysis (solid line) and
the \vWT\ analysis (dashed line).
\label{fig:mumu_fits}}
\end{figure}

Figures~\ref{fig:emu_fits}, \ref{fig:ee_fits}, and \ref{fig:mumu_fits} show the
average weight functions for the full simulation of all three
dilepton channels. Both the kinematic cuts and the additional complexity of
the collider environment further degrade the resolution from that
obtained in parton-level tests. In particular, for top quark masses less than
140 \gevcc, the distributions are distorted significantly by the $H_T$ cut.
This distortion reduces the precision with which a top mass value in
this range can be measured. It does not, however, introduce any bias in our
top mass determination since the
effect of the $H_T$ cut is modeled in the probability distribution
functions used for the mass fits (Sect. \ref{sec-fit}).

The weight distributions become less sharp as the number of
muons in the final state increases, reflecting the relatively poor measurement
of their momenta. This effect is more pronounced for the \vWT\ analysis.
For this reason, and also because the signal to
background ratio is significantly higher for the $e\mu$ channel than for the
$ee$ or $\mu\mu$ channels, it is important to treat the three channels
separately when extracting the top quark mass.

\section{ Mass Fits }
\label{sec-fit}
\subsection{General Procedure}

We estimate the top quark mass by comparing weight functions from Monte Carlo
\ttbar\ samples, generated at different values of the top quark mass,
with the weight functions for the collider data. We use a maximum likelihood
fit to find the value of the top quark mass for which the Monte Carlo
predictions agree best with the data.

\begin{figure}[htpb]
\centerline{\psfig{figure=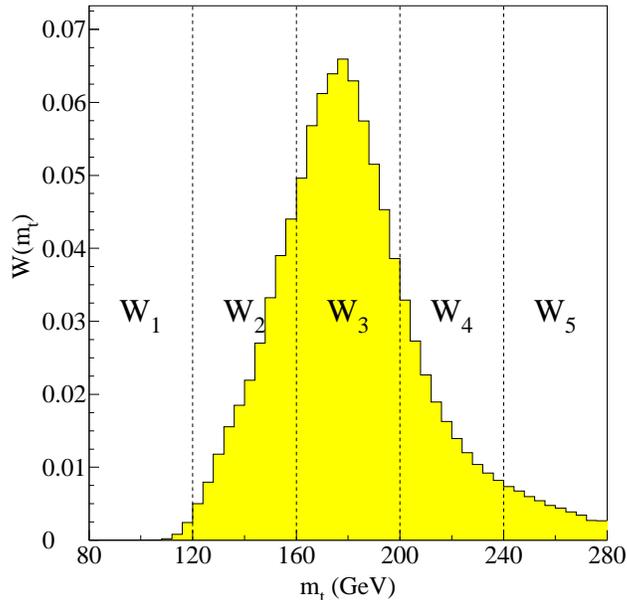,height=3.5 in}}
\caption {The weight function for a typical Monte Carlo event, normalized to
unity. The vertical lines show the five intervals over which the weight
function is integrated.
\label{fig:dilep_wt_bins}}
\end{figure}

For each dilepton event, we compute the weights $W(m_t)$ at 50 values of the
top quark mass between 80 and 280 \gevcc. To fit these 50 values directly we
would need the probability density as a function of 50 arguments, which
is impractical. We can, however, reduce the number of quantities without
losing too much information. The individual weight functions are much
broader than the size of the steps for which the weights are computed.
As shown in Figures \ref{fig:ee_fits}--\ref{fig:mumu_fits}, their rms is 35--40
\gevcc. Therefore, we integrate the weights over five bins 40 \gevcc\ wide, as
shown in Fig.~\ref{fig:dilep_wt_bins}. Since we need information only about the
shape of the weight function, we normalize the area under the function to unity,
such that the integrals over four of the bins are independent quantities.
We thereby reduce the weight function for each event to the four-dimensional
vector
\begin{equation}
\vec{W} = \left(W_1,W_2,W_3,W_4\right),
\end{equation}
where
\begin{eqnarray}
W_1&=&\int_{ 80\ \textrm{GeV}}^{120\ \textrm{GeV}} W(m)\, dm
\end{eqnarray}
and $W_2$, $W_3$, and $W_4$ are computed analogously.

We now maximize the joint likelihood
\begin {eqnarray}
\L &=& {1\over\sqrt{2\pi}\sigma_b}e^{-{\left(n_b-\overline n_b\right)^2
\over2\sigma_b^2}} {\left( n_s + n_b \right)^N e^{-(n_s+n_b)} \over N!} \times
\prod_i^N {n_s f_s(\vec{W}_i|m_t) + n_b f_b(\vec{W}_i) \over n_s + n_b }
\end{eqnarray}
with respect to the parameters $n_s$ (the expected number of signal events),
$n_b$ (the expected number of background events), and $m_t$ (the top quark
mass). The product is taken over all events.
The first term in the likelihood is a Gaussian constraint that forces
the expected number of background events to agree with the background
estimate $\overline n_b$ within its uncertainty $\sigma_b$. The second is a
Poisson constraint that forces the expected number of events to be consistent
with the observed number of dilepton events $N$. The remaining part is the
probability density for the vector $\vec{W}_i$ for the collider data for $n_s$
signal and $n_b$ background events. Here $f_s$ is the probability density
function for signal and $f_b$ for background events. We maximize \L\ with
respect to $n_s$ and $n_b$ at each value of $m_t$ using the {\sc minuit} program
\cite{minuit} to eliminate the nuisance parameters $n_s$ and $n_b$. We are left
with \L\ at the discrete values of $m_t$ for which we have Monte Carlo samples.
Each dilepton channel is treated separately in this fit and the final
likelihood \L\ is the product of the likelihoods from each channel.
We fit a polynomial to $-\ln \L$, the minimum of which gives the measured value
of the top quark mass.

The following sections describe the derivation of the probability density
function for $\vec{W}$, the parametrization of the likelihood functions, and the
fit results.

\subsection{Probability Density Estimation}
\label{sec-pde}

To estimate the continuous functions $f_s$ and $f_b$ from the discrete sample of
Monte Carlo points available for each value of $m_t$ would require a
prohibitively large number of Monte Carlo events to populate the four
dimensional parameter space. We therefore use a probability density
estimation (PDE) technique employing continuous kernels \cite {PDE_paper}.

Consider that each event in the sample is characterized by a set of $d$
uncorrelated values, which are grouped into the $d$-dimensional vector
 $\vec {\zeta}$.  Then the probability density $f$ for any $\vec {\zeta}$ can be
 estimated based on a sample of $\NMC$ Monte Carlo events as
\begin{equation}
\label{eq:pde_def}
f(\vec{\zeta}) = {1 \over \NMC h^d} \sum_{i=1}^{\NMC} K\left( {\vec{\zeta} -
\vec{\zeta}_i \over h}, {\bf C} \right),
\end{equation}
where $\bf C$ is the covariance matrix for the components of $\vec {\zeta}$,
$h$ is a free parameter, and $K$ is the kernel function.

  Any function which is maximal at zero and asymptotically approaches zero as
the absolute value of its argument becomes large would be an acceptable choice
for $K$. For simplicity, we choose a multidimensional Gaussian.
  In our application, the results of applying either the \MWT\ or
\vWT\ techniques to an event is the 4-dimensional vector $\vec {W}$.
The elements of $\vec {W}$ are highly correlated, and so  a
linear transformation must be applied to the data to remove the correlations
before using Eq. \ref{eq:pde_def}:
\begin{equation}
\vec{W}' = {\bf A} \vec{W} \mbox{.}
\end {equation}
  The transformation matrix ${\bf A}$ is chosen so that the covariance matrix
${\bf C}$ of the transformed variables is diagonal.  It can be shown that
for two distinct sources of events (signal and background in our case),
there exists a unique matrix ${\bf A}$ which results in the covariance matrix
for one source to be the identity matrix ${\bf I}$ and that from the other source
to be a general diagonal matrix ${\bf D}$\cite{PDE_paper}.  We choose to have
${\bf C}$ be the identity matrix for background.  The matrix ${\bf A}$ is
computed only once, using the distribution of Monte Carlo \ttbar\ events
generated at all top quark masses.
  After transformation, the kernel function has the form:
\begin {eqnarray}
K\left( {\vec{W}' - \vec{W}_i' \over h } ,{\bf C}\right) =
\prod_{j=1}^d {1 \over \sqrt{2\pi c_j} }
\exp \left( - {\left( {(\vec{W}'-\vec{W}_i')_j/h} \right) ^2
\over 2 c_j} \right)
\end{eqnarray}
where the $c_j$ are the diagonal elements of ${\bf  C}$.

One minor extension of this method is needed to properly model the background.
As described in
Sec. \ref{sec-evt}, the backgrounds in the dilepton channel arise
from a variety of sources.  We assign weight factors $b_j$ such that their
contribution to the probability density corresponds to the relative
strengths of the $n$ background sources:
\begin {equation}
{b_j\NMC_j\over\sum_{i=1}^{n}b_i\NMC_i}={\overline{n}_{j}\over\overline{n}_b},
\end {equation}
where $\NMC_j$ is the number of Monte Carlo events and
$\overline{n}_{j}$ is the number of events expected from the $j^{th}$
background source. The estimate for the probability density for an event
weight vector $\vec{W}$ is then given by:
\begin{equation}
\label{eq:dilep_sig_pdf}
f_s(\vec{W}|m_t) = {1 \over Nh^4} \sum_{i=1}^{N} K\left(
{\vec{W}' - \vec{W}_i' \over h }, {\bf D} \right)
\end{equation}
for signal  and
\begin {eqnarray}
\label{eq:dilep_bkg_pdf}
f_b(\vec{W}) =
{1 \over \left(\sum_{j=1}^{n} b_j \NMC_j\right)h^4}
\sum_{j=1}^{n} b_j \sum_{i=1}^{\NMC_j}
 K\left( {\vec{W}' - \vec{W}_i' \over h }, {\bf I} \right)
\end{eqnarray}
for background.

  The remaining step is to fix the value of the free parameter $h$ to maximize
the expected resolution of the measurement.  Using the ensemble test method
described below, we find that values of $h$ in the range 0.1 -- 0.4 are
preferred, and we choose $h = 0.3$.

\subsection {Ensemble Tests}
Ensemble tests are mock experiments in which the dilepton events are
simulated using a Monte Carlo program with a known top quark mass ($m_t^{\rm
MC}$) and processed in exactly the same manner as the collider data.  The
procedure is as follows:  if there are $N_j$ events in the $j^{th}$ decay
channel, we draw $N_j$ events from the MC samples for this decay channel.
We then select a random number between 0 and 1 for each event. If the random
number is greater than $\overline n_{j} /  N_j $, we take an event from the
signal sample.  Otherwise we select an event from the
background sample. If there are multiple sources of background, another random
number  is selected in order to decide the source of background
from which to draw the event.
We then fit the ensemble using the maximum likelihood procedure described
above.  We repeat this procedure for a large number of ensembles (typically
1000).  In this manner we can gauge the statistical  properties of the maximum
likelihood estimate of the top quark mass, $\widehat {m_t}$.

We characterize the width of the (in general not Gaussian) distribution of fit
results by half the length of the shortest interval in $m_t$ that contains
68.3\%  of the ensembles, $R^{68}$.

\subsection{Parametrization of the Likelihood Function}
\label{sec-like}

We fit a polynomial to the values of $-\ln \L$ computed for different top quark
masses. The fitted top quark mass is the value of $m_t$ for which the
polynomial assumes its minimum $-\ln\L_0$. The statistical uncertainty $\delta
m_t$ due to the finite size of the event sample is given by half of the
interval in $m_t$ for which $-\ln\L<-\ln\L_0+{1\over2}$.

We have a choice of what order polynomial, and how many points
around $\L_0$, to include in the fit.
The values of $\widehat{m_t}$ and $\delta \widehat{m_t}$ returned by the fit
depend on these
choices. We therefore perform ensemble tests to select the choice that gives
the most accurate values. For the fitted top quark mass this means agreement
with the input mass used to generate the ensembles. For the uncertainty it
means agreement with the observed scatter of ensemble results.

We fit quadratic and cubic polynomials to five to eleven
points, centered on the point of maximum likelihood. Table
\ref{tab:dilep_lnl_fits} gives the results of ensemble tests using these
fitting options. The cubic does not improve the accuracy of the fitted mass
and we therefore choose to fit the $-\ln \L$ points with a quadratic polynomial.

\begin{table}
\begin{center}
\caption {Results of ensemble tests using the \vWT\ algorithm showing the
effect of different parametrizations of the $-\ln \L$ function. The fits are
polynomials of degree $m$ to $n$ points.}
\label{tab:dilep_lnl_fits}
\begin {tabular}{cccccccc}
\multicolumn{2}{c}{Fit} & \multicolumn{3}{c}{$m_t^{\rm MC}=150$ \gevcc}
                        & \multicolumn{3}{c}{$m_t^{\rm MC}=200$ \gevcc} \\
$n$ & $m$ & Median & Mean & $R^{68}$ & Median & Mean & $R^{68}$ \\
 & & \gevcc & \gevcc & \gevcc & \gevcc & \gevcc & \gevcc \\
\hline
~5 & 2 & 152.2 & 154.1  & 13.4 & 198.1  & 197.8 & 18.6 \\
~7 & 2 & 151.6 & 154.0  & 13.0 & 198.2  & 198.1 & 19.0 \\
~9 & 2 & 151.9 & 154.5  & 13.6 & 198.8  & 199.4 & 18.9 \\
~9 & 3 & 151.6 & 151.8  & 13.3 & 196.0  & 190.0 & 19.6 \\
11 & 3 & 151.9 & 152.5  & 13.8 & 193.4  & 196.3 & 19.3 \\
\end {tabular}
\end {center}
\end {table}

The width of the fitted quadratic polynomial increases with the number of
points included in the fit. We choose the number of points that results in pull
distributions of unit widths. If $\widehat{m_t}$ is
an unbiased estimate of $m_t^{\rm MC}$ with a Gaussian resolution of width
$\delta \widehat{m_t}$, then the pull
\begin {equation}
s = { \widehat{m_t} - m_t^{\rm MC} \over \delta \widehat{m_t} }
\label{eq:pull_def}
\end {equation}
is normally distributed around zero with unit width. 
We fit Gaussians to histograms of the pulls for all ensembles
generated with the same $m_t^{\rm MC}$.
The pull widths are tabulated
in Table \ref{tab:dilep_pulls_mwt} for the \MWT\ algorithm and in Table
\ref{tab:dilep_pulls} for the \vWT\ algorithm.

\begin{table}
\begin{center}
\caption {Pull means and widths from ensemble tests of the \MWT\ algorithm.}
\label{tab:dilep_pulls_mwt}
\begin {tabular}{ccccr@{.}l}
$m_t^{\rm MC}$ & $n=5$ & $n=7$ &\multicolumn{3}{c}{$n=9$} \\
\gevcc       & Width & Width & Width & \multicolumn{2}{c}{Mean} \\
\hline
    130    &  1.16  &  0.90 &  0.79 &    0&65  \\
    140    &  1.01  &  0.90 &  0.81 &    0&38  \\
    150    &  1.12  &  0.95 &  0.87 &    0&13  \\
    160    &  1.34  &  1.12 &  1.03 &    0&12  \\
    170    &  1.26  &  1.08 &  0.99 &    0&11  \\
    180    &  1.24  &  1.08 &  0.98 &    0&00  \\
    190    &  1.12  &  1.02 &  1.03 & $-$0&06  \\
    200    &  1.17  &  1.10 &  1.06 & $-$0&11  \\
    210    &  1.09  &  1.04 &  1.04 & $-$0&09  \\
\end {tabular}
\end {center}
\end {table}

\begin{table}
\begin{center}
\caption {Pull means and widths from ensemble tests of the \vWT\ algorithm.}
\label{tab:dilep_pulls}
\begin {tabular}{ccccr@{.}l}
$m_t^{\rm MC}$ & $n=5$ & $n=7$ &\multicolumn{3}{c}{$n=9$} \\
\gevcc       & Width & Width & Width & \multicolumn{2}{c}{Mean} \\
\hline
         130   & 1.22  & 1.04  & 1.04 &    0&58 \\
         140   & 1.09  & 0.97  & 0.88 &    0&40 \\
         150   & 1.03  & 0.92  & 0.86 &    0&16 \\
         160   & 1.18  & 0.99  & 0.96 &    0&17 \\
         170   & 1.17  & 1.06  & 0.98 &    0&08 \\
         180   & 1.27  & 1.11  & 1.03 &    0&03 \\
         190   & 1.16  & 1.05  & 0.99 & $-$0&07 \\
         200   & 1.07  & 1.10  & 1.02 & $-$0&08 \\
         210   & 1.08  & 1.01  & 1.03 & $-$0&08 \\
\end {tabular}
\end {center}
\end {table}

\begin{figure}[htpb]
\centerline{\psfig{figure=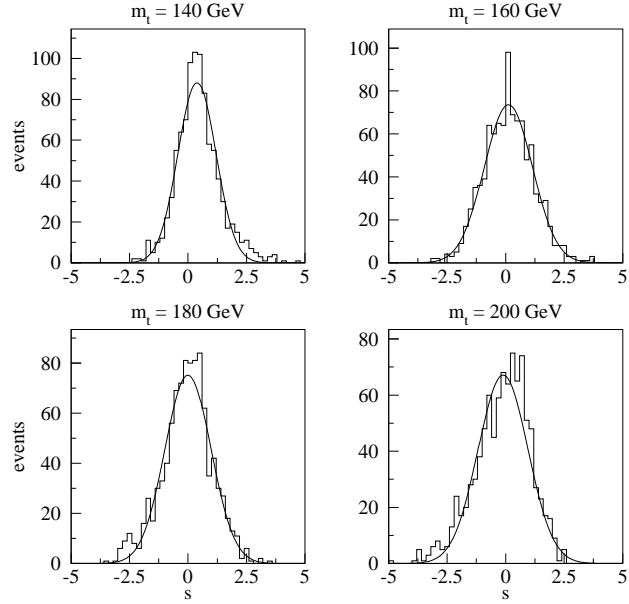,height=3.5 in}}
\caption {Pull distributions for the \MWT\ algorithm. The smooth curves are fits
to Gaussians.}
\label{fig:dilep_pulls_mwt}
\end{figure}

\begin{figure}[htpb]
\centerline{\psfig{figure=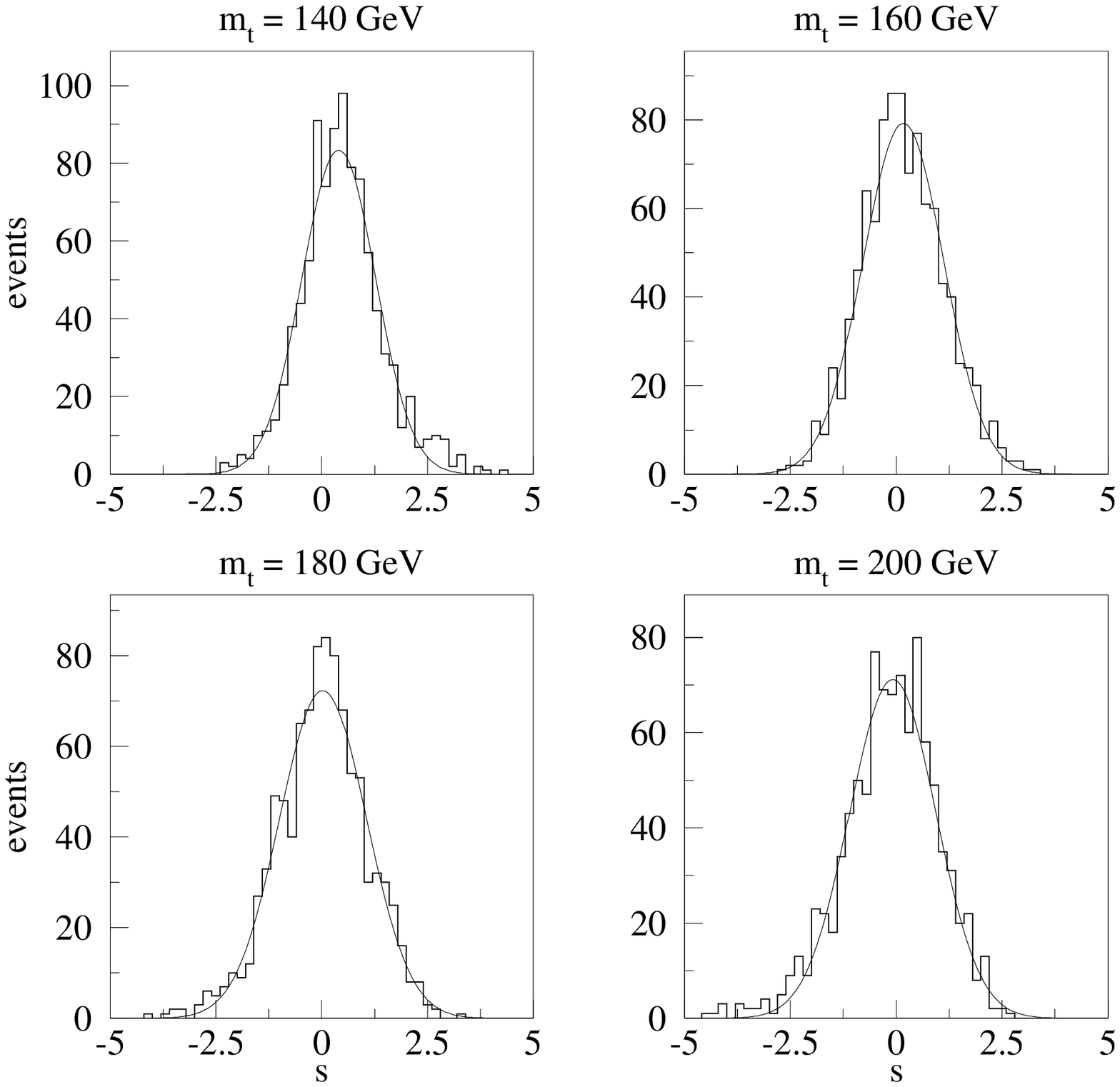,height=3.5 in}}
\caption {Pull distributions for the \vWT\ algorithm. The smooth curves are
fits to Gaussians.}
\label{fig:dilep_pulls}
\end{figure}

The fits that include only five points underestimate
$\delta \widehat{m_t}$. The nine point fits  give pull widths closest to unity over the
whole range of $m_t$. Therefore we choose to fit the quadratic polynomial to
nine points for the final results.
The pull distributions for ensemble  tests at a variety of top quark masses
are shown in Fig.~\ref{fig:dilep_pulls_mwt} for the \MWT\  algorithm and in Fig.
\ref{fig:dilep_pulls} for the \vWT\  algorithm.

\begin{table}
\begin{center}
\caption {Median and mean of the fitted top quark masses and 68\% confidence
intervals from ensemble tests of the \MWT\ algorithm.}
\label{tab:M1_ensems}
\begin {tabular}{cccc}
$m_t^{\rm MC}$ & Median   & Mean     & $R^{68}$ \\
\gevcc       & \gevcc & \gevcc & \gevcc \\ \hline
    130      &   138.1  & 138.3  &  13.6 \\
    140      &   144.6  & 147.1  &  12.7 \\
    150      &   151.6  & 153.4  &  12.8 \\
    160      &   161.6  & 163.9  &  15.8 \\
    170      &   172.2  & 173.7  &  16.7 \\
    180      &   180.5  & 181.0  &  17.3 \\
    190      &   189.5  & 190.5  &  17.8 \\
    200      &   200.3  & 200.1  &  19.5 \\
    210      &   210.0  & 210.9  &  21.4 \\
\end {tabular}
\end {center}
\end {table}

\begin{table}
\begin{center}
\caption {Median and mean of the fitted top quark masses and 68\% confidence
intervals from ensemble tests of the \vWT\ algorithm.}
\label{tab:dilep_mass_tests}
\begin {tabular}{cccc}
$m_t^{\rm MC}$ & Median   & Mean     & $R^{68}$ \\
\gevcc       & \gevcc & \gevcc & \gevcc \\ \hline
    130        &   138.2  & 139.8  &  18.1 \\
    140        &   145.9  & 147.5  &  13.9 \\
    150        &   151.9  & 154.5  &  13.6 \\
    160        &   161.5  & 163.5  &  14.4 \\
    170        &   172.2  & 173.0  &  16.2 \\
    180        &   180.5  & 181.3  &  18.1 \\
    190        &   188.7  & 189.6  &  17.7 \\
    200        &   198.8  & 199.4  &  18.9 \\
    210        &   210.1  & 210.0  &  20.2 \\
\end {tabular}
\end {center}
\end {table}

\begin{figure}[htpb]
\centerline{\psfig{figure=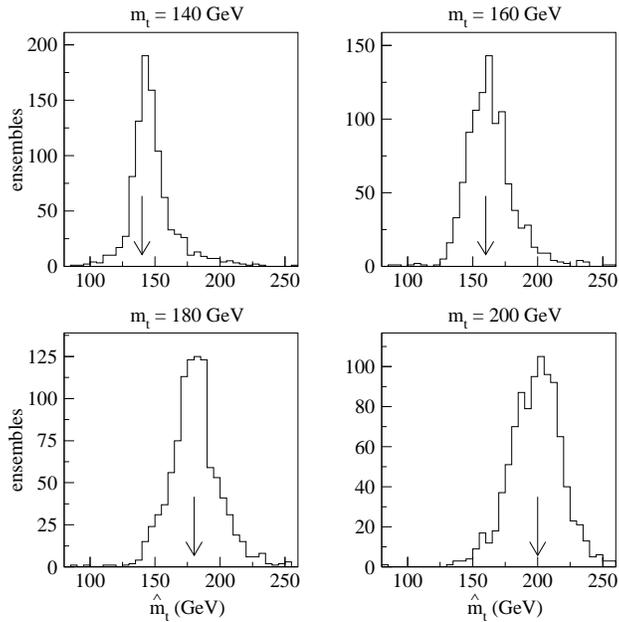,height=3.5 in}}
\caption {Distribution of $\widehat{m_t}$ from ensemble tests of the
\MWT\ algorithm. The arrows point to the input mass.}
\label{fig:dilep_mass_tests_mwt}
\end{figure}

\begin{figure}[htpb]
\centerline{\psfig{figure=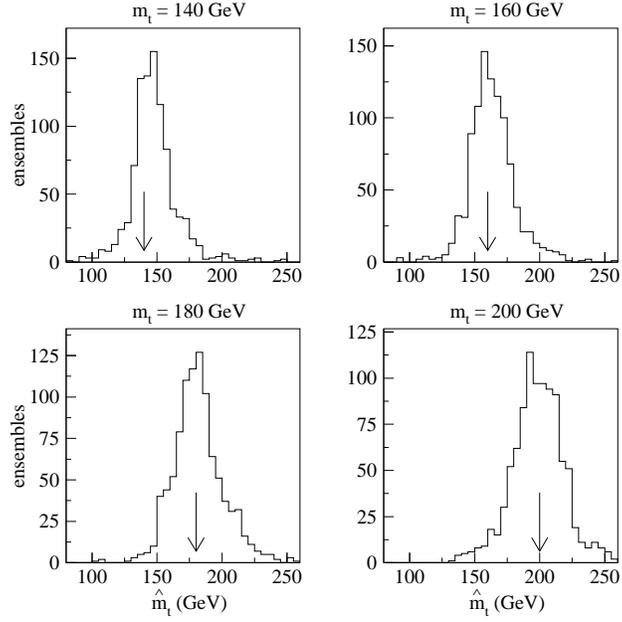,height=3.5 in}}
\caption {Distribution of $\widehat{m_t}$ from ensemble tests of the \vWT\
algorithm. The arrows point to the input mass.}
\label{fig:dilep_mass_tests}
\end{figure}

Tables \ref{tab:M1_ensems} and \ref{tab:dilep_mass_tests} list the median and
mean fitted top quark masses from ensemble tests using a quadratic fit to nine
points.
The differences between $\widehat{m_t}$ and $m_t^{\rm MC}$ at masses below 150
\gevcc\ can be traced to the small number of events available to model some of
the backgrounds ($Z\to\ell\ell$, $WW$). For these background processes the
selection efficiency is so low that a significant increase in the number of
Monte Carlo events that satisfy the selection criteria is not possible due to
limited computing resources.
When we replace these small samples with large samples picked
randomly from a smooth distribution these differences vanish. For fitted
masses above about 150 \gevcc, these differences become small.
We choose not to correct the results for this effect. It is included in the
uncertainty assigned to the fit procedure in Sect. \ref{sec-syst_fit}.
Figures~\ref{fig:dilep_mass_tests_mwt} and \ref{fig:dilep_mass_tests} show that
for the two algorithms, the peak of the $\widehat{m_t}$ distribution is
consistent  with  $m_t^{\rm  MC}$.

\subsection {Results}
\label{sec-fit_results}

\begin{figure}[htpb]
\centerline{\psfig{figure=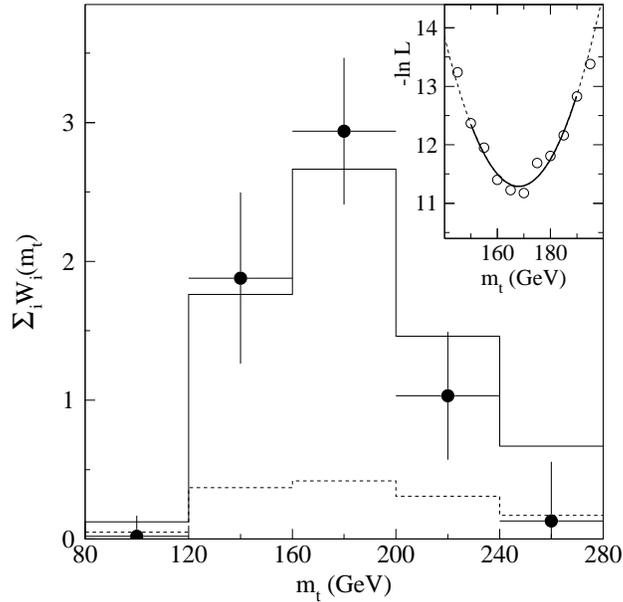,height=3.5 in}}
\caption {
Summed event weight function $\sum_i\vec{W}_i$ for the data sample (points),
the fitted signal plus background (solid), and the background
alone (dashed) for the \MWT\ algorithm. The
error bars indicate the rms observed for five event samples in ensemble
tests. The inset shows the corresponding fit to $-\ln \L$, drawn as a solid line
in the region considered in the fit.}
\label{fig:dilep_result_mwt}
\end{figure}

\begin{figure}[htpb]
\centerline{\psfig{figure=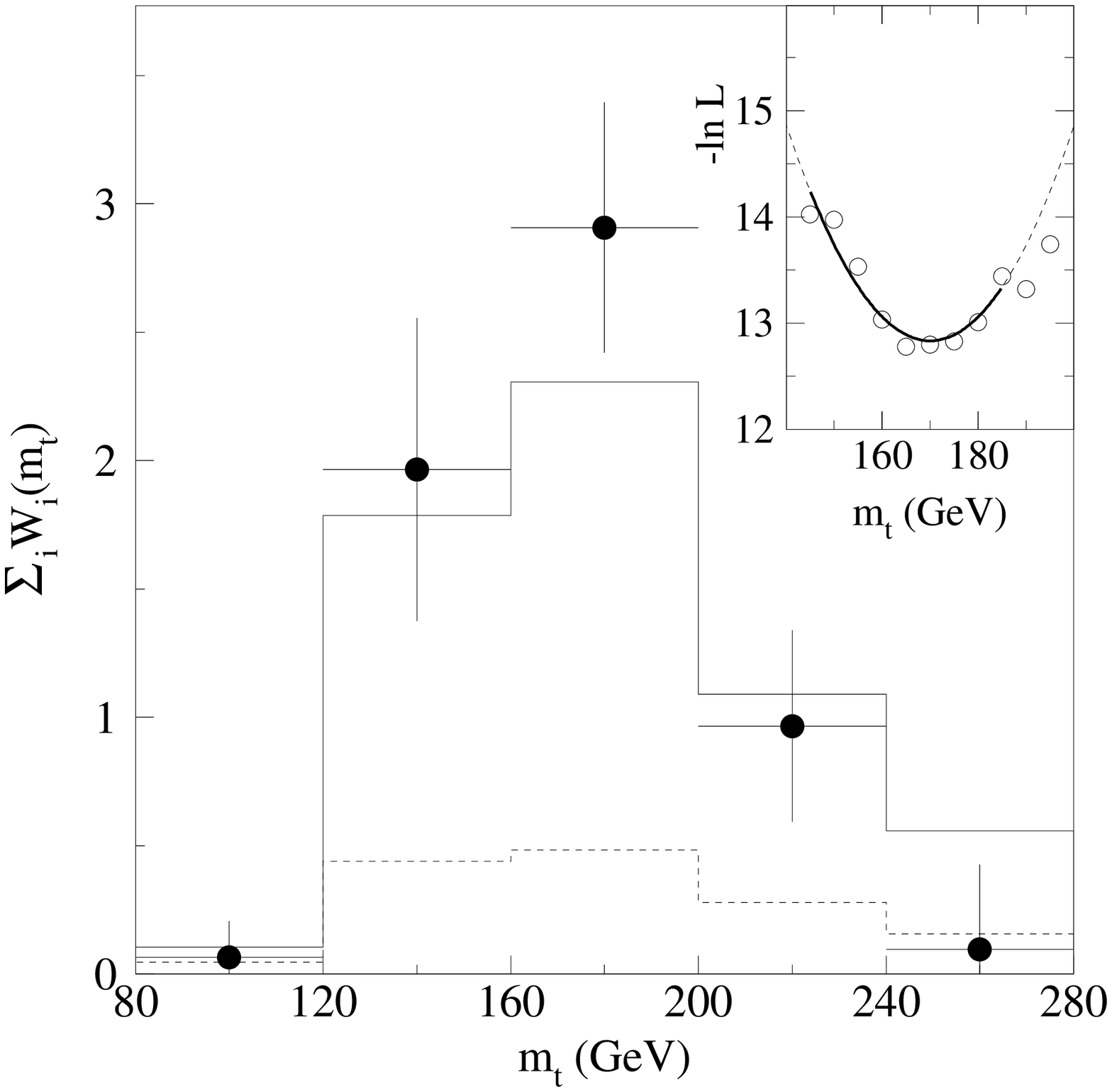,height=3.5 in}}
\caption {
Summed event weight function $\sum_i\vec{W}_i$ for the data sample (points),
the fitted signal plus background (solid),  and the background
alone (dashed) for the \vWT\ algorithm. The
error bars indicate the rms observed for five event samples in ensemble
tests. The inset shows the corresponding fit to $-\ln \L$, drawn as a solid line
in the region considered in the fit.}
\label{fig:dilep_result}
\end{figure}

Applying the procedure outlined above to the dilepton event sample, we find
\begin{equation}
\label{eq:dilep_result_mw}
m_t = 168.2 \pm 12.4 \mbox{ (stat) \gevcc}
\end{equation}
for the \MWT\ algorithm and
\begin{equation}
\label{eq:dilep_result}
m_t = 170.0 \pm 14.8 \mbox{ (stat) \gevcc}
\end{equation}
for the \vWT\ algorithm. Figures~\ref{fig:dilep_result_mwt} and
\ref{fig:dilep_result} compare $\sum_i\vec{W}_i$ for collider data to the
fitted signal plus background shapes. The insets show the corresponding fits to
$-\ln \L$.

\begin{figure}[htpb]
\centerline{\psfig{figure=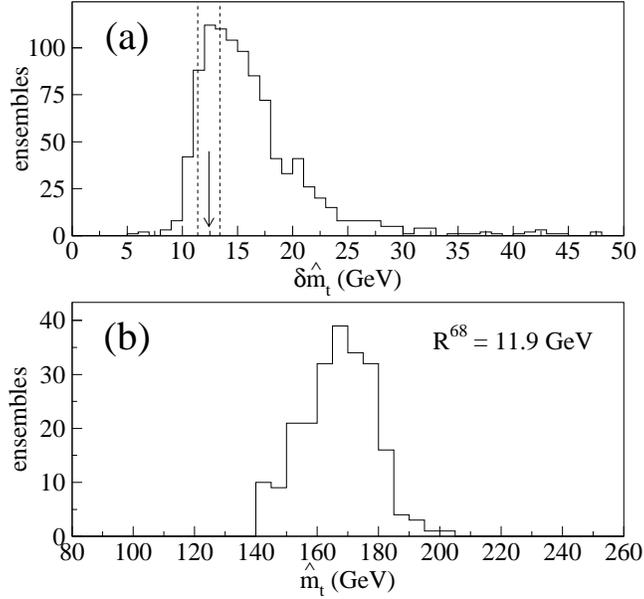,height=3.5 in}}
\caption {
(a) Distribution of uncertainties $\delta \widehat{m_t}$ obtained from ensemble tests for
the \MWT\ algorithm with $m_t^{\rm MC} = 170$ \gevcc. The arrow marks the value
returned by the fit to the data (12.4 \gevcc). (b) Distribution of
$\widehat{m_t}$ for the ensembles with $\delta \widehat{m_t}$ between the dashed lines in
(a).}
\label{fig:dilep_accurate_ens_mwt}
\end{figure}

\begin{figure}[htpb]
\centerline{\psfig{figure=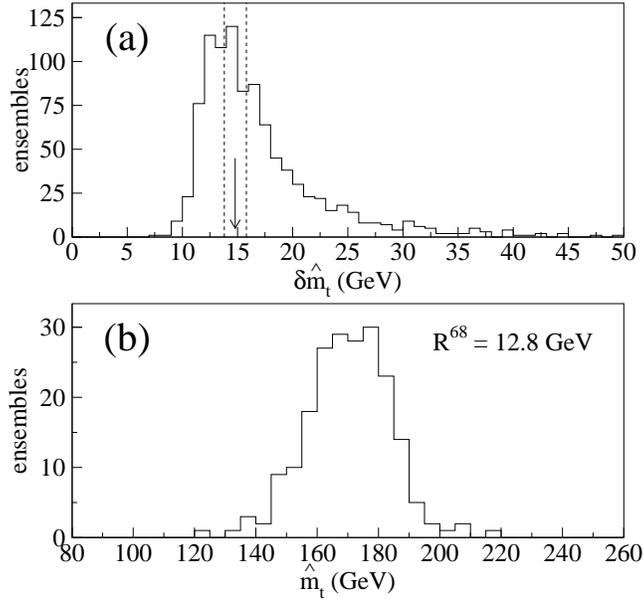,height=3.5 in}}
\caption {
(a) Distribution of uncertainties $\delta \widehat{m_t}$ obtained from ensemble
tests of the \vWT\ algorithm with $m_t^{\rm MC} = 170$ \gevcc. The arrow marks
the value returned by the fit to the data (14.8 \gevcc). (b) Distribution of
$\widehat{m_t}$ for the ensembles with $\delta \widehat{m_t}$ between the
dashed lines in (a).}
\label{fig:dilep_accurate_ens}
\end{figure}

In Figures \ref{fig:dilep_accurate_ens_mwt}(a) and \ref{fig:dilep_accurate_ens}(a)
we compare the statistical uncertainties for the \MWT\ and \vWT\
analyses with the distribution of $R^{68}$ observed in ensemble tests with
$m_t^{\rm MC} = 170$ \gevcc. For the \MWT\ analysis there is a 21\% probability to
obtain a smaller statistical uncertainty than 12.4 \gevcc\ and for the \vWT\
analysis there is a 47\% probability to obtain a smaller statistical uncertainty
than 14.8 \gevcc. The pull distributions indicate that
$\delta\widehat{m_t}$ is a good estimate of the statistical uncertainty. We
verify this by considering the subset of ensembles with $\delta\widehat{m_t}$
consistent with the observed  value.
Figures~\ref{fig:dilep_accurate_ens_mwt}(b) and \ref{fig:dilep_accurate_ens}(b)
show the distribution of mass estimates $\widehat{m_t}$ for the ensembles with
$\delta \widehat{m_t}$ between the dashed lines in (a). The widths $R^{68}$ of
all such
ensembles are consistent with the observed values of $\delta \widehat{m_t}$.

The $e\mu$ channel, with the largest number of events and smallest background,
should dominate the result of the fit, while the $\mu\mu$  channel with only
one event and a sizeable background should have the least effect. We therefore
also fit separately the five events from the $ee$ and $e\mu$ samples and the
three $e\mu$ events. Table  \ref{tab:dilep_fit_vars}
lists the results. This table also shows the effect of varying the degree of
the polynomial used to fit $-\ln \L$ and the number of points included
in the fit. No excursions comparable to the statistical
uncertainty of the measurement are seen in the results of any of these
variations.

\begin{table}
\begin{center}
\caption {Results of several variations of the maximum likelihood fit to the
data. The fits are polynomials of degree $m$ to $n$ points.}
\label{tab:dilep_fit_vars}
\begin {tabular}{cccr@{}lr@{}l}
Channels & \multicolumn{2}{c}{Fit} & \multicolumn{4}{c}{Fitted Mass (\gevcc)} \\
         &        $n$ & $m$ & \multicolumn{2}{c}{\MWT}
& \multicolumn{2}{c}{\vWT}  \\ \hline
$ee,e\mu,\mu\mu$ &        ~5  &  2  & 166&\PM12 & 169&\PM11 \\
                 &        ~7  &  2  & 168&\PM12 & 170&\PM13 \\
                 &        ~9  &  2  & 168&\PM12 & 170&\PM15 \\
                 &        11  &  3  & 167&$^{+11}_{-13}$ & 171&\PM16 \\ \hline
$ee,e\mu$        &        ~5  &  2  & 166&\PM13 & 173&\PM12 \\
                 &        ~7  &  2  & 167&\PM12 & 172&\PM15 \\
                 &        ~9  &  2  & 168&\PM13 & 173&\PM14 \\
                 &        11  &  3  & 166&$^{+11}_{-13}$ & 172&$^{+13}_{-15}$ \\ \hline
$e\mu$           &        ~5  &  2  & 173&\PM15 & 169&\PM14 \\
                 &        ~7  &  2  & 173&\PM13 & 169&\PM13 \\
                 &        ~9  &  2  & 173&\PM13 & 170&\PM15 \\
                 &        11  &  3  & 172&$^{+13}_{-15}$ & 170&$^{+15}_{-16}$ \\
\end {tabular}
\end {center}
\end {table}

\section{ Systematic Errors }
\label{sec-syst}
\subsection {Estimation of Systematic Uncertainties}

Systematic uncertainties give rise to biases in the result of the analysis no
matter how many events are analyzed. They are due to differences between the
collider data and our signal or background models. Variation in the event
selection or the fit procedure, which in general also result in a change in the
final result when applied to a small sample of events, do not represent
systematic uncertainties. Rather, these are statistical effects and are properly
accounted for by our use of a maximum likelihood fit to define the statistical
uncertainty.

Systematic uncertainties can, in general, be estimated using
ensemble tests in which a mismatch is introduced between the conditions under
which the ensembles are created, and the assumptions used in the probability
density estimation. In most cases we vary conditions in the
ensembles and then analyse them with the same probability density functions
used for the collider data, \ie, assuming the nominal conditions.
Any deviation of the fitted mass values from the mass used when generating the
ensembles indicates a systematic effect. Due to the finite number of Monte
Carlo events available, these systematic effects can be estimated with an
uncertainty of about 1 \gevcc.
Table~\ref{tab:dilep_systematics} summarizes the sources of systematic
uncertainties and their estimated magnitudes. The estimated uncertainties
differ insignificantly between the two algorithms so that we use the average
of the uncertainties from both analyses, weighted by the respective
statistical uncertainty in the measured top quark mass, as  
an estimate for both algorithms. The following sections describe
the individual uncertainties in more detail.

\begin {table}
\begin{center}
\caption {Summary of systematic uncertainties for the dilepton mass fits.}
\label{tab:dilep_systematics}
\begin {tabular}{lccc}
 Source                  & \multicolumn{3}{c}{Uncertainty (\gevcc)} \\
                         & \MWT  & \vWT\   & average \\ \hline
 Jet Energy Scale        &  2.0  &   2.9   &   2.4   \\
 Multiple Interactions   &  1.4  &   1.2   &   1.3   \\
 Background Model        &  0.9  &   1.5   &   1.1   \\
 Signal Generator        &  2.3  &   1.1   &   1.8   \\
 Monte Carlo Sample Size &  0.3  &   0.3   &   0.3   \\
 Likelihood Fit          &  0.9  &   1.3   &   1.1   \\
\hline
 Total                   &  3.5  &   3.9   &   3.6   \\
\end {tabular}
\end {center}
\end {table}

\subsection {Jet Energy Scale}
\label{sec-syst_jscale}

To propagate the jet energy scale uncertainty (section
\ref{sec-jscale}) to the top mass
measurement, we generate signal Monte Carlo samples ($m_t = 170\ \gevcc$)
and background samples with jet energy responses one standard deviation
higher
and lower than the nominal response. We also scale the energy in the calorimeter
that is not included in any jet by the same factor as the jets,
and the \mpt\ is recomputed to reflect the scale change. We then create Monte
Carlo ensembles from the scaled samples and fit them using the probability
density functions generated with the nominal jet energy response.
Table~\ref{tab:dilep_jet_scale} shows the results of this mismatch in jet
energy scale. Averaging the upward and downward excursions of the median
results in a systematic uncertainty of 2.0 \gevcc\ for the
\MWT\ algorithm and 2.9 \gevcc\ for the \vWT\ algorithm.

\begin {table}
\begin{center}
\caption {Effect of varying the jet energy response in ensemble tests with
$m_t= 170\ \gevcc$.}
\label{tab:dilep_jet_scale}
\begin {tabular}{ccc}
Jet Scale        & \multicolumn{2}{c}{Median $\widehat{m_t}$ (\gevcc)} \\
                 & \MWT            & \vWT        \\ \hline
$+2.5\%+0.5$ \gev & 172.9           & 174.0          \\
Nominal          & 172.2           & 172.2          \\
$-2.5\%-0.5$ \gev & 168.9           & 168.3          \\
\end {tabular}
\end {center}
\end {table}

\subsection {Signal Monte Carlo Generator}

The accurate determination of the top quark mass depends on the signal Monte
Carlo  providing a faithful description of \ttbar\ events.  Some features, in
particular gluon radiation and parton fragmentation, are  only modeled
approximately by \HERWIG\ and other reasonable approximations exist. In the
absence of large samples of \ttbar\ events, none of them
can be directly excluded.
To test the sensitivity of the result to the Monte
Carlo generator, we generate ensembles of events with the \ISAJET\ event
generator. We simulate the detector response using \GEANT\ and analyse them
in the standard way. We then fit the weight functions of ensembles of these
events with the probability density functions obtained from Monte Carlo events
generated by the \HERWIG\ program. Tables \ref{tab:dilep_gen_sys_mwt} and
\ref{tab:dilep_gen_sys} list the results.
For a given top quark mass, we take the difference $\Delta$Median between the
medians of the results from the \ISAJET\ samples 
(Tables \ref{tab:dilep_gen_sys_mwt} and \ref{tab:dilep_gen_sys}) 
and the \HERWIG\ samples
(Tables \ref{tab:M1_ensems} and \ref{tab:dilep_mass_tests}). We compute the
average of the magnitude of these differences for all top quark masses, 
2.3 \gevcc\ for the \MWT\ algorithm and 1.1 \gevcc\ for the \vWT\ algorithm, and
assign these values as the systematic uncertainty in the top quark mass
measurement.

In addition, we have performed studies to directly assess the impact of gluon
radiation by varying the fraction of events with gluon radiation in a \HERWIG\
Monte Carlo sample by 50\%. This results in a change of 1.3 \gevcc\ in the
measured top quark mass, which is quite consistent with the uncertainties quoted
above based on \HERWIG-\ISAJET\ differences.

\begin {table}
\begin{center}
\caption {Results of analyzing ensembles of events generated by \ISAJET\ with
the \MWT\ algorithm.}
\label{tab:dilep_gen_sys_mwt}
\begin {tabular}{ccccr@{.}lr@{.}l}
$m_t^{\rm MC}$ & Median & Mean & $R^{68}$ 
& \multicolumn{2}{c}{$\Delta$Median} & \multicolumn{2}{c}{$\Delta$Mean} \\
\gevcc & \gevcc & \gevcc & \gevcc & \multicolumn{2}{c}{\gevcc}
& \multicolumn{2}{c}{\gevcc} \\ \hline
    140         & 143.6  & 145.0  & 14.4 & $-$1&0 & $-$2&1 \\
    150         & 151.0  & 151.6  & 14.3 & $-$0&6 & $-$1&8 \\
    160         & 160.0  & 161.4  & 16.4 & $-$1&6 & $-$2&5 \\
    170         & 169.0  & 168.6  & 17.3 & $-$3&2 & $-$5&1 \\
    180         & 178.0  & 178.4  & 18.0 & $-$2&5 & $-$2&6 \\
    190         & 186.2  & 186.9  & 19.8 & $-$3&3 & $-$3&6 \\
    200         & 197.2  & 196.1  & 20.2 & $-$3&1 & $-$4&0 \\
    210         & 206.7  & 206.1  & 22.1 & $-$3&3 & $-$4&8 \\
\end {tabular}
\end {center}
\end {table}

\begin {table}
\begin{center}
\caption {Results of analyzing ensembles of events generated by \ISAJET\ with
the \vWT\ algorithm.}
\label{tab:dilep_gen_sys}
\begin {tabular}{ccccr@{.}lr@{.}l}
$m_t^{\rm MC}$ & Median & Mean & $R^{68}$ 
& \multicolumn{2}{c}{$\Delta$Median} & \multicolumn{2}{c}{$\Delta$Mean} \\
 \gevcc &  \gevcc &  \gevcc &  \gevcc &  \multicolumn{2}{c}{\gevcc}
& \multicolumn{2}{c}{\gevcc} \\ \hline
    140         & 145.9  & 147.8  & 15.6 &    0&0 &    0&3 \\
    150         & 152.6  & 154.4  & 15.4 &    0&7 & $-$0&1 \\
    160         & 160.1  & 161.6  & 15.8 & $-$1&4 & $-$1&9 \\
    170         & 170.8  & 171.6  & 17.6 & $-$1&4 & $-$1&4 \\
    180         & 179.1  & 179.5  & 18.2 & $-$1&4 & $-$1&8 \\
    190         & 189.4  & 188.7  & 18.5 &    0&7 & $-$0&9 \\
    200         & 198.6  & 198.3  & 19.5 & $-$0&2 & $-$1&1 \\
    210         & 206.8  & 205.6  & 20.3 & $-$3&3 & $-$4&4 \\
\end {tabular}
\end {center}
\end {table}

We studied the sensitivity of the results to variations in our choice of parton
distribution functions. We expect the sensitivity to parton distribution
functions to be larger for the \MWT\ analysis because it uses them explicitly in
the mass reconstruction. Our default choice is the CTEQ3M set of parton
distribution
functions \cite{CTEQ3M}. We also perform ensemble tests with weight functions
derived using MRSA$'$ parton distribution functions \cite{MRSA} with
three different values of $\Lambda_{\rm QCD}$. The Monte Carlo
events for the ensembles were generated with an input mass of 170 \gevcc\ and
CTEQ3M parton distribution functions in the generation and the top mass
reconstruction. The results are summarized in Table~\ref{tab:pdf}. The variation
in the median of the ensemble tests is 20 MeV. We conclude that any sensitivity
to parton distribution functions is negligible compared to other systematic
effects in the generation of the Monte Carlo samples.

\begin {table}
\begin{center}
\caption {Results of varying the choice of parton distribution functions (pdf)
in the \MWT\ analysis.}
\label{tab:pdf}

\begin {tabular}{lcc}
pdf                                   & Median & Mean \\
                                      & \gevcc & \gevcc \\ \hline
CTEQ3M                                & 172.25 & 173.67   \\
MRSA$'$ ($\Lambda_{\rm QCD}=266$ MeV) & 172.27 & 173.66   \\
MRSA$'$ ($\Lambda_{\rm QCD}=344$ MeV) & 172.27 & 173.51   \\
MRSA$'$ ($\Lambda_{\rm QCD}=435$ MeV) & 172.26 & 173.38   \\
\end {tabular}
\end {center}
\end {table}

\subsection {Background Shape}

The modeling of the background also depends on a Monte Carlo simulation.  In
addition,  for some sources of background ($Z\to\ell\ell$, $WW$) very few
Monte Carlo events satisfy the selection criteria.
To estimate how sensitive the result is to the poorly constrained distribution
of these events, we use dummy models instead of the Monte Carlo samples. These
models assume that the $W(m_t)$  distributions for these
backgrounds are Gaussian, with a width chosen
randomly between 20 and 60 \gevcc.  In one of the models (``low mass''), the
mean of the Gaussian was randomly selected between 120 and 160 \gevcc, and in
the other  (``high mass'') between 180 and 220 \gevcc. We then perform
ensemble tests using the known background components plus the dummies to
estimate the background probability  densities, with events drawn from the
standard signal and background models.  The results are listed in Table
\ref{tab:dilep_bkg_shape}. Based on the
observed shifts in the median $\widehat{m_t}$ the uncertainties are 0.9 \gevcc\
and 1.5 \gevcc\ for the \MWT\ and \vWT\ analyses, respectively.

\begin {table}
\begin{center}
\caption {Effect of introducing dummy models for the poorly modeled 
portion of the background.}
\label{tab:dilep_bkg_shape}
\begin {tabular}{ccc}
Background Model & \multicolumn{2}{c}{Median $\widehat{m_t}$ (\gevcc)} \\
                 & \MWT           & \vWT        \\ \hline
  Low Mass  &  172.9 &  172.7 \\
  Nominal   &  172.2 &  172.2 \\
  High Mass &  172.0 &  171.2 \\
\end {tabular}
\end {center}
\end {table}

\subsection {Multiple Interactions}

The beams in the Tevatron are structured into six proton and six antiproton
bunches. Proton and antiproton bunches collide every 3.5 $\mu$s in the center
of the detector. More than one \ppbar\ interaction can take place during a
crossing and the detector sees the superposition of all these interactions. At
the mean luminosity at which the data were taken
($7.5\times10^{30}/\hbox{cm}^2/\hbox{s}$) on average 1.3 interactions occur
per crossing. Since the cross section for the production of high-$p_T$
secondaries is small, it is very unlikely that more than one of these
interactions produces high-$p_T$ particles or jets. However,
the Monte Carlo models do not include the effect of the additional low-$p_T$
particles due to multiple interactions during the same crossing.

 There are two ways in which these additional interactions may affect
the reconstructed event.  First,  the additional particles deposit energy in the
calorimeter, some of which falls into the jet cones. Second, the additional
tracks  may confuse the algorithm that determines the $z$-position of the
interaction vertex, leading to
mismeasurement of the jet directions. The jet energy scale calibration
accounts for the former effect on average.
To study the latter effect, we add particles from one or two simulated
additional \ppbar\ interactions to a sample of 5000 Monte Carlo \ttbar\ decays
with $m_t=170\ \gevcc$. The signatures of the resulting events in the detector
are simulated by the \GEANT\ program. The events are reconstructed by the
same programs as the collider data.
For this study ensemble tests are of little help, since the small sample
sizes prohibit the generation  of a large number of independent ensembles. We
estimate the size of the systematic effect by comparing
the $W(m_t)$ distributions in the samples with zero, one, and two additional
interactions.  Although the  resolution of
the $z$ vertex degrades with the additional interaction, the effect on the
$W(m_t)$  distribution is modest. The difference in mean
between a sample without additional interactions and the sample in
which 33\% of the events have one and 36\% two additional interactions,
approximating the conditions at which the data were taken, is only 0.6 \gevcc\ 
for the $\nu$WT analysis. A change of this
magnitude is roughly equivalent to a change of 1.2 \gevcc\ in the top quark
mass. For the \M WT analysis we get a similar value, 1.4 \gevcc.

\subsection {Likelihood Fit and Monte Carlo Statistics}
\label{sec-syst_fit}

There are systematic uncertainties in the value of
the top quark mass that minimizes $-\ln \L$. These arise both from the finite
number of Monte Carlo events used in  determining the $-\ln \L$ points and the
choice of function to fit these points.

To estimate the effect of the Monte Carlo sample size, we split the signal
Monte Carlo samples into five subsets and repeat the fit to the data using
each subset as the signal model.  The rms variation observed in the central
value is then divided by $\sqrt{5}$, yielding a systematic uncertainty of 0.3
\gevcc\ for either algorithm.

To estimate the uncertainty arising from the choice of the parabolic fit to
nine likelihood points, we fit Monte Carlo ensembles with  $m_t= 170\ \gevcc$
using a variety of parametrizations and observe the resulting changes
in the median of $\widehat m_t$. We fit quadratic polynomials to five and seven
points
and cubic polynomials to nine and eleven points. The largest variations
of 0.9 \gevcc\ (\MWT) and 1.3 \gevcc\ (\vWT) give
estimates of the systematic uncertainties.

\section{ Results }
\label{sec-results}
\subsection {Combination of the \MWT\ and \vWT\ Measurements}

The two algorithms we use give consistent results.
The weights computed by the \MWT\ and \vWT\ algorithms are based on different
aspects of \ttbar\ production and decay and are therefore not completely
correlated. To gauge the degree of correlation, we fit ensembles of \ttbar\
Monte Carlo events for a top quark mass of 170 \gevcc\ using both algorithms.
We then select the subset of these ensembles with likelihood functions of
similar widths as observed in the data (\ie\ those for which the \MWT\
analysis yields  $11.4 <\delta m_t< 13.4$ \gevcc\ and the \vWT\ analysis
yields $13.8 <\delta m_t< 15.8$  \gevcc). Based on  these tests we find that the
correlation  coefficient between the  \MWT\ and \vWT\ algorithms is
0.77. A statistical combination of the results from the two algorithms then
yields
\begin {equation}
m_t = 168.4 \pm 12.3 \mbox{ (stat)} \pm 3.6 \mbox{ (syst) \gevcc}.
\end {equation}
The systematic uncertainties are taken as completely correlated between the two
algorithms. Since they differ insignificantly between the two algorithms we
quote the mean from Table \ref{tab:dilep_systematics}.

\subsection {Combination of the Dilepton and Lepton+Jets Measurements}

The value of the top quark mass obtained from the dilepton channel 
is in good agreement with that found by fitting
$\ttbar\to\l+\hbox{jets}$ events \cite{D0_mtop_lj}, supporting the hypothesis
that
both are due to the decays of the same pair-produced particles. We obtain our
best measurement of the mass of the top quark by combining the results of the
analyses in the two channels. Since the two measurements are statistically
independent the combination is straight forward. The systematic uncertainties
in the combined
measurement are evaluated by propagating the uncertainties in each channel with
correlation coefficients of either 0 (for MC statistics, likelihood fit, and 
background model)
or 1 (for jet energy scale, multiple interactions, and \HERWIG--\ISAJET\
differences). We obtain
\begin{equation}
m_t  = 172.1 \pm 5.2 \mbox{ (stat)} \pm 4.9 \mbox{ (syst)}\ \gevcc.
\label{eq:dilep_final_result_all}
\end{equation}
The effective correlation coefficient between the two measurements is 0.15. If
we neglected
all correlations the result would change by less than 200 MeV.

\subsection {Conclusions}

We have reported the measurement of the top quark mass using six dilepton
events. We use maximum likelihood fits to the dynamics of the decays to achieve
maximum
sensitivity to the mass of the top quark. We developed two algorithms for the
computation of the likelihood that exploit complementary features of \ttbar\
production and decay. Both result in very similar measurements of the top quark
mass. They also agree well with the mass measured from fits to
$\ttbar\to\l+\hbox{jets}$ events, supporting the hypothesis that both channels
correspond to decays of the same particle. We combine the mass
measurements from both channels to obtain
\begin {equation}
m_t = 172.1 \pm  7.1\ \gevcc.
\end {equation}

\section*{ Acknowledgements }

%
We thank the staffs at Fermilab and collaborating institutions for their
contributions to this work, and acknowledge support from the 
Department of Energy and National Science Foundation (U.S.A.),  
Commissariat  \` a L'Energie Atomique (France), 
Ministry for Science and Technology and Ministry for Atomic 
   Energy (Russia),
CAPES and CNPq (Brazil),
Departments of Atomic Energy and Science and Education (India),
Colciencias (Colombia),
CONACyT (Mexico),
Ministry of Education and KOSEF (Korea),
and CONICET and UBACyT (Argentina).

\end{document}
